\documentclass[a4paper, aps, pra, twocolumn, longbibliography]{revtex4-2}  

\usepackage{amssymb}
\usepackage{amsmath}
\usepackage{graphicx}
\usepackage{bbm}
\usepackage{bbold}
\usepackage{color}
\usepackage{comment}
\usepackage{lipsum}
\usepackage{tikz}
\usepackage[normalem]{ulem}
\usetikzlibrary{arrows}
\makeatletter
\newcommand\@erelb@r[1]{%
	\mathrel{\tikz[baseline=-.5ex]\draw[#1] (0,0)--(0.2,0);}
}
\newcommand{\erelbar}[1]{\@erelbar#1}
\def\@erelbar#1#2{%
	\ifcase\numexpr#1*4+#2\relax
	\@erelb@r{-}\or     
	\@erelb@r{->}\or    
	\@erelb@r{-|}\or    
	\@erelb@r{->|}\or   
	\@erelb@r{<-}\or    
	\@erelb@r{<->}\or   
	\@erelb@r{<-|}\or   
	\@erelb@r{<->}\or   
	\@erelb@r{|-}\or    
	\@erelb@r{|->}\or   
	\@erelb@r{|-|}\or   
	\@erelb@r{|<->|}\or 
	\@erelb@r{|<-}\or   
	\@erelb@r{|<->}\or  
	\@erelb@r{|<-|}\or  
	\@erelb@r{|<->|}    
	\else
	\@wrong
	\fi
}
\makeatother
\definecolor{blue}{rgb}{0,0,1}
\definecolor{grey}{rgb}{0.6,0.6,0.6}

\begin{document}
	
\title{Understanding Dissipative Transport of Bosons and Fermions from Growing Interfaces}

\title{A Unified Interface Model for Dissipative Transport of Bosons and Fermions}


\author{Y. Minoguchi$^{\,1*}$, J. Huber$^{\,1*}$, L. Garbe$^{\,1}$, A. Gambassi$^{\,2}$, P. Rabl$^{\,1,3,4,5}$}
\affiliation{$^1$Vienna Center for Quantum Science and Technology, Atominstitut, TU Wien, 1020 Vienna, Austria}
\affiliation{$^2$SISSA – International School for Advanced Studies and INFN, via Bonomea 265, 34136, Trieste, Italy}
\affiliation{$^3$Technical University of Munich, TUM School of Natural Sciences, 85748 Garching, Germany}
\affiliation{$^4$Walther-Mei\ss ner-Institut, Bayerische Akademie der Wissenschaften, 85748 Garching, Germany}
\affiliation{$^5$Munich Center for Quantum Science and Technology (MCQST), 80799 Munich, Germany} 

\date{\today}

\begin{abstract}
	We study the directed transport of bosons along a one dimensional lattice in a dissipative setting, where the hopping is only facilitated by coupling to a Markovian reservoir.
	By combining numerical simulations with a field-theoretic analysis, we investigate the current fluctuations for this process and determine its asymptotic behavior.  
	These findings demonstrate that dissipative bosonic transport belongs to the KPZ universality class and therefore, in spite of the drastic difference in the underlying particle statistics, it features the same coarse grained behavior as the corresponding asymmetric simple exclusion process (ASEP) for fermions. 
	However, crucial differences between the two processes emerge when focusing on the full counting statistics of current fluctuations.  
	By mapping both models to the physics of fluctuating interfaces, we find that dissipative transport of bosons and fermions can be understood as surface growth and erosion processes, respectively. 
	Within this unified description, both the similarities and discrepancies between the full counting statistics of the transport are reconciled.  
	Beyond purely theoretical interest, these findings are relevant for experiments with cold atoms or long-lived quasi-particles in nanophotonic lattices, where such transport scenarios can be realized.
\end{abstract}

\maketitle

Transport is one of the most common non-equilibrium processes and thus the subject of intense research in many fields of physics \cite{Nazarov09_book,Kohler05_review,Lepri03_review,Hartnoll15,Beenakker91_review,Chou11_review_bio,Bertini15_RMP}.
Of considerable interest in this context is the analysis of elementary paradigmatic models, which capture essential aspects of transport in terms of simplified abstractions of broader classes of physical processes. 
A prominent example in statistical mechanics is the asymmetric simple exclusion process (ASEP) \cite{Spitzer91_ASEP, Derrida98_ASEP_Review,Golinelli06ASEP_Review,Kriecherbauer10pedestrian,Mallick15ASEP_Review}, which describes random asymmetric hopping of hard-core particles or fermions on a one-dimensional lattice. 
This model has been the subject of intense research in statistical physics and many exact results led to a deeper understanding of the physics far from equilibrium \cite{Rost81ASEP2surf,Gwa92,Johansson2000,Praehofer02}. 
Most important for this work, it has been shown that the ASEP can be mapped on a model of a randomly fluctuating interface \cite{Rost81ASEP2surf} whose coarse grained properties are well described in terms of the celebrated Kardar–Parisi–Zhang (KPZ) equation \cite{Kardar86}. This establishes an interesting connection between these two seemingly unrelated fields.
\begin{figure}
	\centering
	\includegraphics[width=\linewidth]{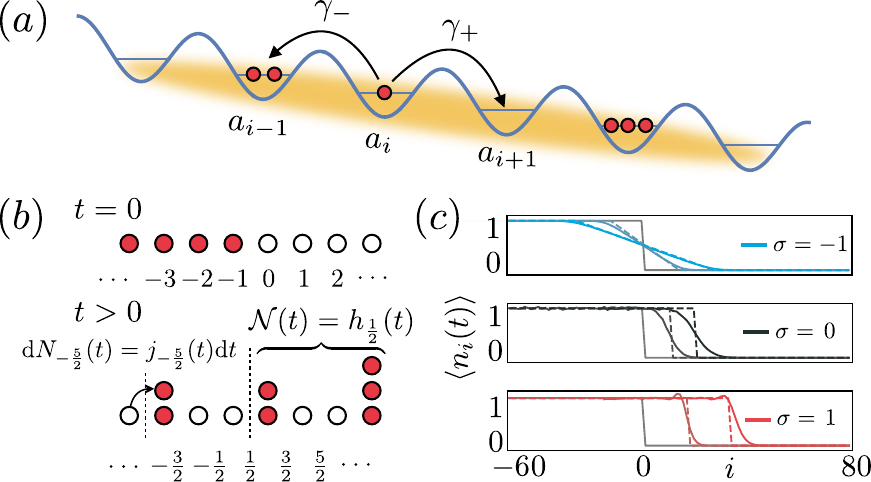}
	\caption{
		(a) Sketch of the dissipative transport setup, where bosons hop along a tilted 1D lattice. 
        Due to the coupling to a Markovian, low-temperature reservoir (yellow cloud), hopping becomes incoherent with asymmetric rates $\gamma_+$ and $\gamma_-$. 
		(b) To characterize transport, the lattice is initially prepared with one particle in each lattice site $i\leq0$. In this case $\mathcal{N}(t)$ denotes the total number of particles to the right of this boundary at a later time $t$. 
		The increments ${\rm d}N_{i+\frac{1}{2}}=j_{i+\frac{1}{2}}{\rm d}t$ count how many particles hop over the dual site $i+\frac{1}{2}$ per unit time. 
		 (c) Comparison of the subsequent evolution of the average site occupation numbers $\langle n_i\rangle$ for fermions ($\sigma=-1$), bosons ($\sigma=+1$) and the ZRP ($\sigma=0$), for $\gamma_+=\gamma$ and $\gamma_-=0$. The dashed line corresponds to the analytical solution of the (noiseless) Burgers equation.}
	\label{fig:1}
\end{figure}

In this Letter, we study the dissipative transport of bosonic particles along a one-dimensional lattice, which can be described by an asymmetric simple \emph{inclusion} process (ASIP) \cite{Giardina07,Giardina10,Bernard18_ASIP,Garbe23,Note0}.
Physically, this model is motivated by recent experimental setups with cold atoms in optical lattices~\cite{Krinner15_observation,Labouvie16_bistability,Gou20_tunable,Scherg21_observing,Huang22_intrinsic} and with long-lived photonic, polaritonic and plasmonic excitations in nanophotonic structures~\cite{Klaers10_bose,Marelic16_spatiotemporal,Schmitt16_spontaneous,Deng02_condensation,Kasprzak06_bose,Deng10_exciton,Daskalakis15_spatial,Baboux16_bosonic,Lerario17_room,Fontaine22_kardar,Hakala18_bose}, where such bosonic transport can be realized. In contrast to Pauli's exclusion principle, which motivates the ASEP for fermions, bosonic atoms or quasi-particles favor the hopping to neighboring sites that are already occupied. 
This property gives rise to an opposite, bosonically-enhanced \textit{inclusive} transport, which drastically changes the overall particle flow and can even lead to condensation phenomena~\cite{Grosskinsky11condensation,Cao14dynamics_ASIP_condensation,Garbe23} that are naturally forbidden for fermions.

Here we investigate the dissipative transport of bosonic particles, starting from a step initial condition where all bosons completely occupy half of the infinite lattice. 
Surprisingly, this analysis reveals that the transport of bosons belongs to the KPZ universality class.
Surprisingly, in spite of their fundamentally different exchange statistics, fluctuations of dissipative currents of both bosons and fermions exhibit the same scaling behavior.
However, more refined numerical investigations show that bosonic and fermionic currents are distributed according to different types of Tracy-Widom distributions (TWD) \cite{Tracy94GUE,Tracy96GOE_GSE}, which characterise the distribution of the largest eigenvalue of a random matrix ensemble.
In particular we find that fermionic currents are distributed according the TWD of the Gaussian unitary ensemble (GUE) while for bosons we find the Gaussian orthogonal ensemble (GOE).
Finally, we discuss the physical origin of this subtle but important difference and provide a unified picture for the two processes by mapping them to the physics of fluctuating interfaces.
In this picture and for the considered initial conditions, bosonic (fermionic) transport corresponds to random growth (erosion) which leads to an interface without (with) curvature.
According to the refined universality conjecture \cite{Praehofer00PNG}, the difference in the interface geometry then explains the different counting statistics of the transport.

\emph{Model.}---We consider a one dimensional (1D) lattice as shown in Fig.~\ref{fig:1}(a), where bosonic atoms or other quasi-particles hop incoherently between neighboring sites with asymmetric rates $\gamma_+$ and $\gamma_-$. 
This is the case, for example, for a reservoir-assisted tunneling processes in the presence of an external bias, and we refer to Refs.~\cite{Haga21,Garbe23} for possible experimental realizations.
We model the dynamics of this system by a quantum master equation for the density operator $\rho$,
\begin{equation} \label{eq:ME_quantum}
	\dot \rho= \gamma_+ \sum_{i=1}^{L-1} \mathcal{D}[a_{i+1}^\dag a_i] \rho +  \gamma_- \sum_{i=2}^L \mathcal{D}[a_{i-1}^\dag a_i] \rho,
\end{equation} 
where $L$ is the total number of lattice sites, the $a_i$ $(a_i^\dag)$ are bosonic or fermionic annihilation (creation) operators at site $i$ and $\mathcal{D}[A]\rho = A \rho A^\dagger - \frac{1}{2}\{ A^\dagger A, \rho \}$. 
For simplicity, we focus on the totally asymmetric case, $\gamma_+=\gamma$ and $\gamma_-=0$; as long as $\gamma_+ \ne \gamma_-$, none of the following predictions crucially depend on this simplification. 

In the absence of Hamiltonian dynamics, the density operator $\rho$ evolving under Eq.~\eqref{eq:ME_quantum} remains diagonal in the Fock basis $\vert \vec{n}= (n_1,\ldots, n_L)\rangle$, where $n_i = 0,1,2,\ldots$ is the number of bosons on site $i$. 
It is then sufficient to consider the dynamics of  the diagonal elements $P_t(\vec{n})=\langle \vec{n} \vert \rho(t)\vert \vec{n}\rangle$, which obey
\begin{equation} \label{eq:ME_class}
	\partial_t P_t(\vec{n}) 
	= 
	\gamma \sum_{i=1}^{L-1}  \left[T_i^{+}T_{i+1}^{-} - 1\right] n_i(1+\sigma n_{i+1})P_t(\vec{n}).
\end{equation}
Here, we introduced the operators $T_i^{\pm}f(\vec{n}) = f(\ldots,n_{i-1},n_i\pm 1 ,n_{i+1},\ldots)$ as well as the parameter $\sigma_\in\{0,\pm 1\}$ for the particle statistics. 
The latter allows us to compare the bosonic ASIP $(\sigma=+1)$ with the fermionic ASEP $(\sigma=-1)$ and with a zero-range process (ZRP) $(\sigma = 0)$ for independent classical particles, for which the hopping rate does not depend on the occupation of the target site \cite{Evans05_review_ZRP,Evans14}.

The dynamics of the probability distribution $P_t(\vec{n})$ in Eq.~\eqref{eq:ME_class} can be sampled numerically by a Monte Carlo simulation of the individual site occupation numbers $n_i(t)$, which are updated at each time step $\mathrm{d}t$ according to 
\begin{equation} \label{eq:ni}
	n_i(t+\mathrm{d}t) = n_i(t) + \mathrm{d}N_{i-\frac{1}{2}}(t) - \mathrm{d}N_{i+{\frac{1}{2}}}(t).
\end{equation}
Here, $\mathrm{d}N_{i-\frac{1}{2}}=0,1$ are Poisson processes that indicate a particle hopping onto site $i$ by jumping over the dual site $i-\frac{1}{2}$, placed in between $i-1$ and $i$ [and similarly $\mathrm{d}N_{i+\frac{1}{2}}$ encodes the jumps between sites $i$ and $i+1$; see Fig.~\ref{fig:1}(b)].
The probabilities for these events to occur are $\mathrm{Pr}[\mathrm{d}N_{i+{\frac{1}{2}}}(t)=1]= \Gamma_{i\rightarrow i+1}\mathrm{d}t$, where the hopping rates $ \Gamma_{i\rightarrow i+1}=\gamma n_i(1+\sigma n_{i+1})$ depend on the particle statistics $\sigma$. 
Note that Eq.~\eqref{eq:ni} has the form of a continuity equation, from which the current $j_{i+\frac{1}{2}} = \mathrm{d}N_{i+\frac{1}{2}}/\mathrm{d}t$ can be directly read off. 

\emph{Transport properties.}---In the following we consider the initial configuration depicted in Fig.~\ref{fig:1} (b), where at $t=0$ the left half of the lattice is filled with one particle per site i.e. $n_{i\le 0} = 1$, while the sites to the right are empty.
By numerically evolving Eq.~\eqref{eq:ni} and averaging over $10^3$ realizations, we obtain the mean density profiles shown in Fig.~\ref{fig:1}(c). 
This plot shows that bosons propagate like a \textit{shockwave}, which is accelerated compared to the diffusive advective transport of the ZRP and the even slower dynamics of fermions, which spread like a \textit{rarefaction} wave~\cite{SI}.

\begin{figure}
	\centering
	\includegraphics[width=1\linewidth]{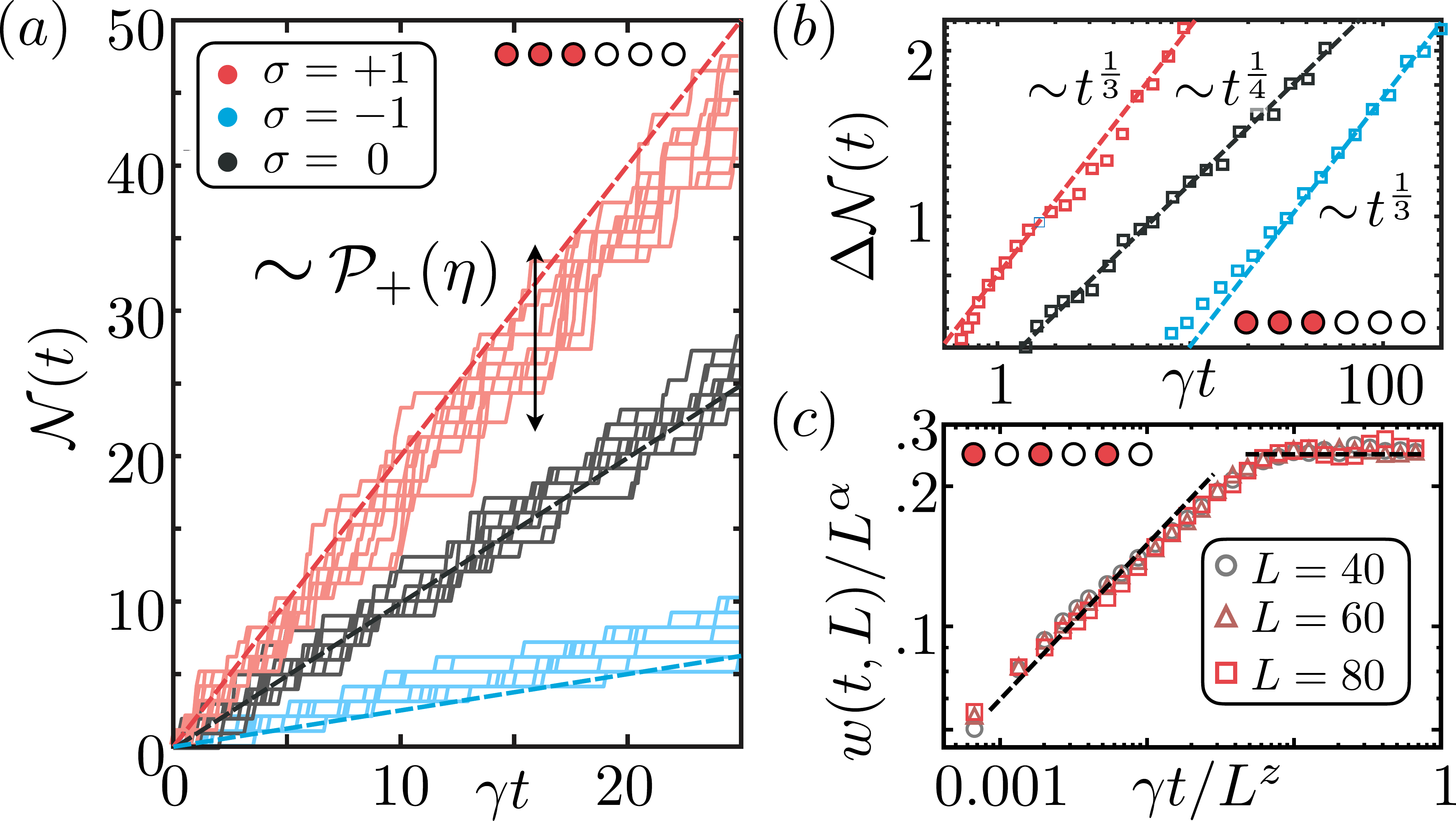}
	\caption{(a) Individual realizations of the total number of transported particles, $\mathcal{N}(t)$, for different statistics $\sigma$ and step initial conditions.
		The thick lines indicate $\langle\mathcal{N}(t)\rangle =j_\sigma t$, as predicted by the noiseless Burgers equation.
		The corresponding currents are $j_+=2\gamma$ (bosons), $j_0=\gamma$ (ZRP) and $j_-=\gamma/4$ (fermions). 
        The fluctuations grow with $t^\beta$ and so do the deviations from the noiseless Burgers equation due to the finite mean of the TWD. 
		(b) Scaling of $\Delta \mathcal{N}(t)$ as a function of time. 
		(c) Scaling collapse of $w(t,L)$ for ASIP. For this plot, we took a lattice of size $L$ and periodic boundary conditions and uniform initial condition with every other site occupied. 
	}\label{fig:2}
\end{figure}

To go beyond averaged quantities, we now take a closer look at individual trajectories, which represent the outcomes of different experimental realizations of the dynamics. Specifically, we consider the total number of transported particles, $\mathcal{N}(t)=\sum_{i>0}n_i(t)$, which is equal to the integrated current through a boundary at the origin [see dashed line in Fig.~\ref{fig:1}(b)].  
We show $\mathcal{N}(t)$ for a few realizations of the dynamics in Fig.~\ref{fig:2}(a). Overall, these trajectories increase roughly linearly in time, with $\langle \mathcal{N}(t)\rangle \simeq j_{\sigma} t$, and an average current $j_\sigma$ that depends on the particle statistics. The shot to shot fluctuations around this trend, $\Delta \mathcal{N}^2 (t)= \langle \mathcal{N}^2(t)\rangle - \langle \mathcal{N}(t)\rangle^2$, are shown in Fig.~\ref{fig:2}(b) with a fit of the form $\Delta \mathcal{N}(t)\sim t^\beta$.
For the ZRP the fluctuations increase with $\beta \simeq 1/4$, while a clearly distinct scaling exponent $\beta \simeq 1/3$ is observed for both fermions and bosons. 
Accordingly, in spite of their drastically different microscopic behavior reflected in the mean transport, these simulations indicate that bosons exhibit the same non-trivial scaling of fluctuations as fermions \cite{Beijeren85,Johansson2000}.

\emph{Fluctuating hydrodynamics.}---
To explain these numerical findings, we consider the hydrodynamic limit $n_i(t)\rightarrow n(t,x)$, where we approximate the bosonic occupation numbers for the individual sites by a continuous density $n(t,x)$, coarse-grained over many adjacent lattice sites and time steps $\mathrm{d}t$. 
Applying either the Fokker-Planck approximation on Eq.~\eqref{eq:ME_class} or the framework of nonlinear fluctuating hydrodynamics \cite{vanBeijeren12_NLFHD,Spohn14NLFHD},
we derive in \cite{SI} an approximate stochastic differential equation for $n(t,x)$, which turns out to be the \textit{stochastic Burgers equation} \cite{Forster77}
\begin{equation}\label{eq:SBE}
	\partial_t n + \gamma (1+2\sigma n)\partial_x n = \nu \partial_x^2 n -\partial_x\xi,
\end{equation}
with a viscosity $\nu =\gamma/2$ and a white noise $\xi \equiv \xi(t,x)$ with $\langle \xi(t,x)\xi(t^\prime,x^\prime) \rangle =  D\delta(x-x^\prime)\delta(t-t^\prime)$ and $D = 2\nu n(x,t)[1+\sigma n(x,t)]$. 
Since the precise magnitude of the noise does not alter the scaling of the current fluctuations, at long times we replace $n(t,x) \simeq \bar{n}$ by the steady state filling $D\simeq 2\nu \bar{n}(1+\sigma\bar{n}) $ in what follows.
For $\sigma = -1$,  Eq.~\eqref{eq:SBE} is then a well established effective description of the ASEP \cite{Beijeren85,Bertini97ASEP_KPZ,Wijland01}.
By omitting the viscosity, we obtain a deterministic equation, which for the case of bosons $(\sigma = +1)$ indeed predicts a shockwave traveling at a bosonically-enhanced current $ j_{+} =2\gamma$ (see Fig.~1(c) and \cite{SI}).
For the ZRP $(\sigma = 0)$, we have a linear transport equation with a current $j_0 = \gamma $ that is only half of that for bosons.
In case of fermions $(\sigma = -1)$, the analysis in Refs. \cite{Kriecherbauer10pedestrian,Krapivsky10_book_kinetic,SI} predicts a rarefaction wave, which matches the density profile shown in Fig.~\ref{fig:1}(c) with a mean current of $j_- = \gamma/4$, eight times smaller than the bosonic current.

\emph{KPZ universality.}---Instead of characterizing the particle dynamics via the local site occupations $n_i$, we may consider the number of particles transported across the dual lattice site, $h_{i+\frac{1}{2}}(t)= \int_0^t\mathrm{d}\tau j_{i+{\frac{1}{2}}}(\tau)$ [see Fig.~\ref{fig:1}(b)]. 
The number of particles on site $i$ is then the difference of particles hopping on and off site $i$ up to time $t$, i.e., $n_i(t) = h_{i-\frac{1}{2}}(t) - h_{i+\frac{1}{2}}(t)$. 
In this representation we identify $\mathcal{N}(t)\equiv h_{\frac{1}{2}}(t)$ and, in the hydrodynamic limit, $h_{i+\frac{1}{2}}(t)\rightarrow h(t,x)$ we obtain the continuity equation
$-\partial_x h(t,x)= n(t,x)$.
Inserting this relation into Eq.~\eqref{eq:SBE} yields
\begin{equation} \label{eq:KPZ_transp}
	\partial_t h =v_{\sigma} - c_\sigma \partial_x  h+ \gamma \sigma (\partial_x h)^2 + \nu \partial_x^2 h + \xi, 
\end{equation}
with  $c_\sigma = \gamma$ and $v_\sigma = 0$, which we introduced for later convenience.
Up to a Galilean transformation, this is the well known KPZ equation \cite{Kardar86, Bertini97ASEP_KPZ}. 
Note that the particle statistics $\sigma$ determines the sign of the KPZ nonlinearity. 
The physics of this equation is characterized by the scaling exponent $z$, which governs the temporal growth of the correlation length $L_{\rm corr} \sim t^{1/z}$ and $\alpha$ characterizes the spatial roughness of the transport fluctuations. Scaling theory then suggests that $\Delta h(t,0) = \Delta \mathcal{N}(t) \sim t^{\alpha/z}$, or $\beta=\alpha/z$ in the notation above (see \cite{SI} for details).
In 1D the scaling exponents $z = \frac{3}{2}$ and $\alpha = \frac{1}{2}$ are known and independent of the sign $\sigma$ of the nonlinearity \cite{SI}.  
This explains the identical exponents $\beta=1/3$ obtained in numerical simulations for both fermions and bosons~\cite{FN}. 
In contrast, for the Edwards-Wilkinson (EW) equation \cite{Edwards82} obtained for the ZRP $(\sigma=0)$, $\alpha$ remains the same, but the dynamical scaling changes to $z = 2$, consistent with the numerically extracted exponent $\beta=1/4$.
To investigate scaling with the system size, we perform additional numerical simulations of the ASIP on a periodic chain, with translationally invariant initial conditions, i.e., with an alternating filling from site to site.
In Fig.~\ref{fig:2}(c) a scaling collapse of $w(t,L)= \langle (L^{-1}\sum_i (h_i(t)-\bar{h}(t))^2)\rangle^{1/2}$, with $\bar{h}(t) = L^{-1}\sum_{i}h_i(t)$, is shown for the theoretically predicted values of $\alpha$ and $z$, with Family-Viczek scaling $w(t,L) = L^{\alpha}f(t/L^z)$ \cite{Family85,Barabasi95boook}. 
This plot then fully confirms that the ASIP belongs to the KPZ universality class and shows that bosonic and fermionic transport are much more similar than expected from the dynamics of averaged quantities. 
This is the first main finding of this work.

\emph{Full counting statistics.}---Let us now go beyond the scaling of second moments and take a closer look at the statistics of $\mathcal{N}(t)$, which at long times is written in the asymptotic form \cite{Johansson2000}
\begin{equation}\label{eq:h_FCS}
	\mathcal{N}(t) = j_{\sigma}t + (\Gamma_\sigma t)^{\beta}\eta.
\end{equation}
Here, the first term describes average linear growth with the current $j_{\sigma}$, predicted by the noiseless Burgers equation.
The second term, instead describes the fluctuations, and from it we have already extracted the time dependent scaling factor $\Gamma_\sigma = \gamma [\bar{n}_\sigma (1+\sigma \bar{n}_\sigma)]^{2}C_\sigma$ where $\bar{n}_\sigma$ corresponds to the long time local density at the origin [see Fig.~\ref{fig:1}(c)].
For fermions with $\bar{n}_- = \frac{1}{2}$ in the origin we find $C_- = 1$ which agrees with the exact solution \cite{Johansson2000}, while for bosons with $\bar{n}_+ = 1$, our simulations suggest $C_+ \simeq 0.3 $  \cite{SI}. 
The random variable $\eta$ is distributed according to the \emph{time-independent} probability distribution $\mathcal{P}_\sigma(\eta)$, which determines the full counting statistics (FCS) of $\mathcal{N}(t)$.

\begin{figure}
	\centering
	\includegraphics[width=1\linewidth]{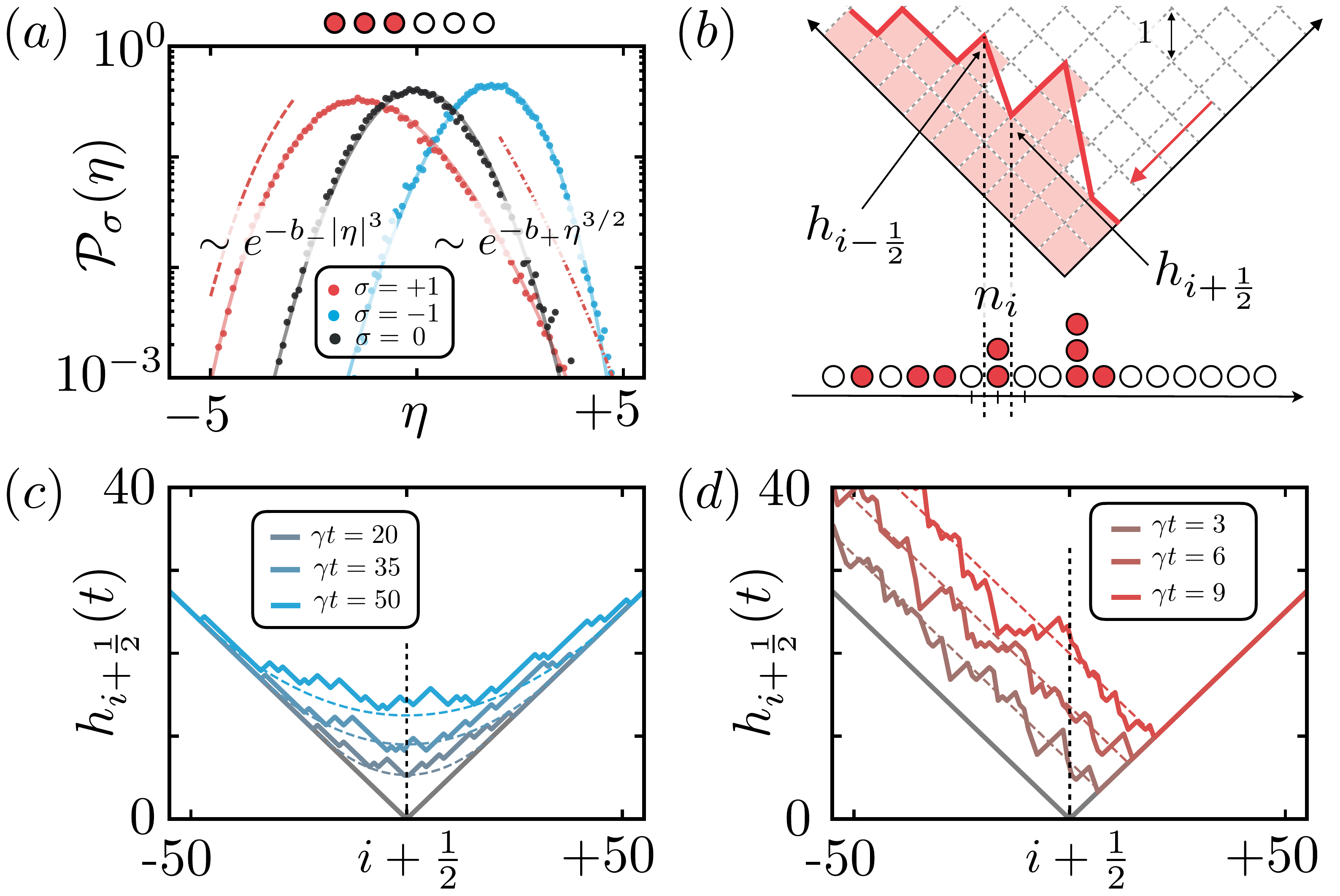}
	\caption{
		(a) The numerically obtained FCS (dots) for the bosons (red), fermions (blue) and ZRP (black) for a step initial condition. 
		The solid lines corresponds to ${\rm TW}_{\rm GOE}(\eta)$ (red), ${\rm TW}_{\rm GUE}(-\eta)$ (blue) and Gaussian (black).
		(b) Mapping the bosonic transport to corner growth. 
		The red arrow corresponds to the direction of particles deposition leading to the growth of the surface height $h_{i+\frac{1}{2}}$ (thick red line).
		Particle transport (solid lines) for fermions (c) and bosons (d) is plotted in terms of $h_i(t)$ for a single realization. 
		The colored dashed lines correspond to the noise and noiseless solution of Eq.~\eqref{eq:KPZ_transp}.
		The height along the black dashed line corresponds to $\mathcal{N}(t)$.
        In the fermionic case (c), the surface develops a curvature, whereas in the bosonic case (d) it stays flat.}\label{fig:3}
\end{figure}

Starting from the step initial conditions, the distribution of $\mathcal{N}(t)$ can be extracted numerically for arbitrary $\sigma$.
For the ZRP, the fluctuations are Gaussian due to the linearity of the EW equation. 
For the case $\sigma = -1$ of fermions, we find the \textit{sign-reversed} TWD of the GUE i.e., $\mathcal{P}_-(\eta) = \mathrm{TW}_{\rm GUE}(-\eta)$ [see Fig.~\ref{fig:3}(a)] in agreement with the exact solution \cite{Johansson2000}. Surprisingly, for  the case of bosons, $\sigma = +1$, and with the \textit{same} initial conditions, we find the TWD of the GOE \textit{without} sign reversal i.e., $\mathcal{P}_+(\eta) = \mathrm{TW}_{\rm GOE}(\eta)$ [see Fig.~\ref{fig:3}(a)].
Accordingly, the FCS of the bosons is sign-reversed and is slightly broader (GOE) compared to its fermionic counterpart (GUE).
This is the second main finding of our work.
Note that the negative (positive) first moment of the GOE (GUE) TWD implies that bosons (fermions) flow slightly slower (faster) than predicted by noiseless hydrodynamics [see Fig.~\ref{fig:2}(a)].

\emph{Large-deviations analysis for bosons.}---It is possible to obtain a physical intuition for the emergence of asymmetric tails in the FCS of the bosons. Let us consider rare events, where up to time $t$ either no particle, $\mathcal{N}(t)\sim 0$, or an excess number of particles, $\mathcal{N}(t)\sim j_{\rm exc}  t$ with $j_{\rm exc}\gg j_+$, were transported. 
In both cases we find from Eq.~\eqref{eq:h_FCS} the scaling $\eta \sim t^{2/3}$. 

With this in mind we start our discussion by considering the right tail $\mathcal{P}_+(\eta \rightarrow \infty)$, i.e., the extreme case of an unusually large number of particles being transported. 
We assume that these events are dominated by shock configurations with a large number of particles $\bar{n}_{\rm s} \gg 1$ concentrated in the front.
This gives rise to an excess current $j_{\rm exc} = \gamma\bar{n}_{\rm s}(1+\bar{n}_{\rm s})$.
The probability that in a single time step $\Delta t$, on hydrodynamics scales, a number of $j_{\rm exc} \Delta t$ particles is transported to the right is $p_{1}$.
Since the shape of the shock front does not change, this probability remains constant and assuming independence of this excess transport at different time step we obtain ${\rm Pr}[\mathcal{N}(t)= j_{\rm exc} t] \sim (p_{1})^{t/\Delta t} \sim \exp(-a_+ t)$ where the unknown constant $a_+$ depends on microscopic details. 
Going from $t$ to the variable $\eta$ we obtain the asymptotic form $P_+(\eta \rightarrow \infty) \sim \exp (-b_+ \eta^{3/2})$ [see red dashed dotted line in Fig.~\ref{fig:3}(a)].

The left tail $\mathcal{P}_-(\eta \rightarrow -\infty)$ is instead obtained, by focusing on rare events where no particle is transported across the origin.
Note that due to the ongoing transport up to the origin and their bosonic nature, particles pile up like $n_0(t) \sim j_+ t$ at the origin, and the probability of \text{not} jumping further right is progressively suppressed with the occupation $p(n_0) \simeq 1- n_0 \gamma \Delta t\simeq (p_{0})^{n_0}$ with $p_0 = 1-\gamma \Delta t$.
Then the probability that for long times $t$ no particle hops to the right can be estimated as $\mathrm{Pr}[\mathcal{N}(t) = 0] \sim p_0 ^1 p_0^2\cdot  \ldots \cdot p_0^{t/\Delta t} \sim \exp(\sum_{i=1}^{t/\Delta t} i ) \sim \exp(- a_- t^2)$. Going again from $t$ to $\eta$ we obtain $\mathcal{P}_+(\eta \rightarrow \infty) \sim \exp(-b_-\eta^3)$ [red dashed line in Fig.~\ref{fig:3}(a)].

While these arguments are too crude to approximate the correct $b_{\pm}$, they capture the correct scaling of the FCS tails, and show that bosonic statistics has a decisive effect on the shape (sign reversal) of the distribution.
They should also be viewed in comparison to the case of fermions or the ASEP where similar arguments for the sign reversed asymptotics were obtained \cite{Krapivsky10_book_kinetic}.

\emph{Unified interface model.}---
In a final step we show how the differences and similarities between bosonic and fermionic transport can be reconciled by mapping both processes onto a common model of a fluctuating interface.
Above, we showed how the transport process can be described in terms of a field $h_{i+\frac{1}{2}}$, which increases by $1$ if a particle jumps from $i$ to $i+1$. 
They key step is to interpret this quantity as the \textit{height} of a randomly increasing interface.
To make connection with the exact results for the ASEP \cite{Johansson2000}, we exploit the freedom in the initial conditions of the height field and set $h_{i-\frac{1}{2}}(t) - h_{i+\frac{1}{2}}(t) = n_i(t) - \frac{1}{2}$. With this convention, every particle configuration can be mapped to a height configuration of this effective interface model [see Fig.~\ref{fig:3}(b) for bosons and \cite{Rost81ASEP2surf,Johansson2000} for fermions] with an initial profile in the shape of a \textit{corner}, $h_{i+\frac{1}{2}}(t=0) = \vert i \vert /2$. 
The number of particles transported to the right corresponds to the interface height in the origin, $\mathcal{N}(t) = h_{\frac{1}{2}}(t)$.

In the hydrodynamic limit the dynamics of the interface height are described by the KPZ equation~\eqref{eq:KPZ_transp} with \textit{drive} $v_\sigma = \frac{\gamma}{2}(1+\frac{\sigma}{2})$ and advection $c_\sigma = \gamma(1+\sigma)$.
Formally we obtain this by setting $h \rightarrow h + \frac{x}{2}$ in Eq.~\eqref{eq:KPZ_transp} (see \cite{SI} for details).
In the case of bosons, there is an additional advection term $c_+$, which is absent for fermions.
This term is responsible for the deposition of particles under an angle parallel to the right wedge [see red arrow in Fig.~\ref{fig:3}(b)] leading to an accumulation of particles on the left side only. Finally, the positive (negative) signs of the KPZ nonlinearity translate into what we shall refer to as \textit{surface growth} (\textit{surface erosion}).
When combined, these three mechanisms result in two very distinct growth patterns, which are depicted in Fig.~\ref{fig:3}(c) and~\ref{fig:3}(d) for fermions and bosons, respectively.

Based on this interface model, the differences in the FCS of bosonic and fermionic transport can now be understood as follows. 
First, neglecting $v_{\sigma}$ and $c_{\sigma}$ for now, the transformation $h\rightarrow - h$ either flips the sign of the KPZ nonlinearity, from growth to erosion $\gamma \rightarrow - \gamma$, or maps bosons $\sigma = 1$ into fermions $\sigma = -1$ and vice versa.
The transmutation of bosons into fermions under this transformation explains why the FCS are, up to a broadening, sign flipped with respect to each other [see Fig.~\ref{fig:3}(a)].
Secondly, we invoke the \textit{refined} classification of the KPZ universality to explain the relative broadening of the FCS between bosons and fermions [see Fig.~\ref{fig:3}(a)]. This refined classification, which was originally conjectured by Pr{\"a}hofer and Spohn~\cite{Praehofer00PNG} and  confirmed both theoretically and experimentally for many different examples,
states that the FCS of interfaces which are curved (flat) are of the GUE \cite{Johansson2000,Praehofer00PNG,Sasamoto10,Dotsenko10,Calabrese10,Takeuchi11exp} (GOE \cite{Borodin07,Praehofer00PNG,Calabrese11,Takeuchi11exp}) type, respectively.
As discussed earlier, while the sign of the KPZ nonlinearity does not affect the scaling exponents, 
our findings suggest that the curvature is very much affected. 
This can already be anticipated from the KPZ equation in Eq.~\eqref{eq:KPZ_transp} in the absence of noise [dashed line in Fig.~3(c-d)].  
For the case of surface growth (erosion) the effective interface corresponding to bosonic (fermionic) transport stays asymptotically flat (curved) around $x = 0$, which explains the different TWD observed in $\mathcal{N}(t)= h(t,x=0)$. 
This is our third main finding. 

\emph{Conclusion.}---
In summary, we have analyzed the ASIP describing the directed dissipative transport of bosonic particles along a 1D lattice. 
Despite their drastically different behavior at the level of noiseless hydrodynamics, we have found surprising similarities between fermionic (ASEP) and bosonic (ASIP) transport in the scaling of the fluctuations. 
However, these two cases map onto two different interface processes, a curved, eroding surface for fermions on the one hand and a flat, growing surface for bosons on the other hand.
This explains the subtle difference in the full counting statistics of the transport.
These fundamental properties, which are exclusively attributed to the particle statistics, can be directly probed, i.e., in cold-atom experiments, where techniques to track the dynamics of individual particles are already available.

\emph{Acknowledgements.}---We thank L. Canet, M.Zündel, F. Helluin, J. Dubail, M. Pr{\"a}hofer and H. Spohn for stimulating discussions. 
This work was supported by the Austrian Science Fund (FWF) through Grant No. P32299 (PHONED) and No. M3214 (ASYMM-LM), and the European Union’s Horizon 2020 research and innovation programme under Grant Agreement No. 899354 (SuperQuLAN). A.G. acknowlegdes financial support from from the PNRR MUR project No. PE0000023-NQSTI. This research
is part of the Munich Quantum Valley, which is supported by the Bavarian state government with funds from the Hightech Agenda Bayern Plus.

\clearpage

\setcounter{equation}{0}
\setcounter{figure}{0}
\setcounter{table}{0}
\makeatletter
\renewcommand{\theequation}{S\arabic{equation}}
\renewcommand{\thefigure}{S\arabic{figure}}
\renewcommand{\bibnumfmt}[1]{[S#1]}
\renewcommand{\citenumfont}[1]{S#1}

\widetext
\begin{center}
	\textbf{ \large Supplementary Material:}
	\textbf{ \large A Unified Interface Model for Dissipative Transport of Bosons and Fermions}
\end{center}

\setcounter{equation}{0} \setcounter{figure}{0} \setcounter{table}{0}
\makeatletter \global\long\def\theequation{S\arabic{equation}}
\global\long\def\thefigure{S\arabic{figure}}
\global\long\def\bibnumfmt#1{[S#1]}
\global\long\def\citenumfont#1{S#1}

\section{Supplement to paragraph ``Fluctuating Hydrodynamics"}

In this section we motivate the effective description of dissipative transport via the stochastic Burgers equation (SBE) given in Eq.~\eqref{eq:SBE} in the main text.
We will do this by two approaches.
In Sec.~\ref{SI:NLFH}, we will present a derivation using the framework of \textit{nonlinear fluctuating hydrodynamics} (NLFHD) \cite{vanBeijeren12_NLFHD_SI}; in Sec.~\ref{SI:FPE_approx}, we will use the \textit{Fokker-Planck approximation} instead.
In Sec.~\ref{SI:Nt}, we then go on to discuss the solutions of the noiseless Burgers equation, which gives the predictions for first moments of $\langle \mathcal{N}(t)\rangle$ (shown as dashed lines in Fig.~\ref{fig:2}(a) in the main text).

\subsection{SBE from Non-Linear Fluctuating Hydrodynamics}
\label{SI:NLFH}

\subsubsection{Conserved Quantity}
We start by considering an equation of motion for the local particle numbers $n_i$ starting from Eq.~\eqref{eq:ME_class}, which has the form of a continuity equation
\begin{equation}
	\langle \dot{n}_i(t)\rangle
	=
	-\partial_x^- \langle J_{i+\frac{1}{2}}(t)\rangle,
\end{equation}
where we defined the backward discrete derivative $\partial_x^- f_i = f_i -f_{i-1} $ and
the current
\begin{equation}
	\langle J_{i+\frac{1}{2}} \rangle = \gamma \langle n_i(1+\sigma n_{i+1})\rangle.
\end{equation}
It turns out that these are the only conserved quantities.
The main quantity of interest in the main paper is $\mathcal{N}(t) = \sum_{i>0} n_i(t)$ and is a sum of these conserved quantities. 
Our goal will be to obtain an effective (stochastic) equation of motion for the \textit{hydrodynamic} variables 
\begin{equation} \label{eq:hydro_avg}
	n(t = j\cdot\tau, x = k\cdot \ell ) 
	= 
	\frac{1}{\ell}\sum_{i = -\ell/2}^{\ell/2-1} \frac{1}{\ell \tau}
	\sum_{m = -\ell/2}^{\ell/2-1}
	n_{k\cdot\ell +i}(j\cdot \tau +m \mathrm{d}t),
\end{equation}
which is a coarse graining over $\ell \gg 1$ adjacent sites of the microscopic mode occupations $n_i$ and over a time $\tau = \ell \gamma^{-1}$ where $\gamma$ is the characteristic time scale of the underlying stochastic dynamics and $\mathrm{d}t = (\ell \gamma)^{-1}$
Usually in the NLFH setting the dynamics are deterministic while the intial states are probabilistic, but here we are in the reverse situation where we start from a deterministic initial state i.e. step initial state which evolves under a stochastic time evolution Eq.~\eqref{eq:ni}.

\subsubsection{Steady State} \label{sec:SS}
For nonlinear systems the average in Eq.~\eqref{eq:hydro_avg} cannot be performed exactly.
In the following we will consider an alternative approach using the steady state $P_{0,\sigma}(\vec{n})$ of Eq.~\eqref{eq:ME_class}, which can be obtained exactly.
We will construct this steady state explicitly for bosons on a ring (periodic boundary conditions) and refer for the case of fermions to \cite{Kriecherbauer10pedestrian_SI,Derrida98_ASEP_Review_SI}.
The steady state is the constant measure $P_{0,+}(\vec{n}) = Z_+^{-1}$ which can be verified by plugging this into Eq.~\eqref{eq:ME_class}
and noticing that the telescoping sum cancels out on a ring.
Therefore we obtain the constant steady state
\begin{equation}\label{eq:P0}
	P_{0,\sigma}(\vec{n}) = \frac{1}{Z_{\sigma}}
	\qquad \text{where}\qquad
	Z_{\sigma} =
	\begin{cases}
		\binom{L + N - 1}{N}& \sigma = +1,0 \\
		\binom{L}{N} & \sigma = -1 
	\end{cases}
\end{equation}
is the normalization keeping track of all possible configurations of $N$ particles on $L$ sites.
This distribution can be rewritten as a product (measure) over the distribution over all sites
\begin{equation}
	P_{0,\sigma}(\vec{n}) = \prod_{i=1}^{L} p_{0,\sigma}(n_i),
\end{equation}
and hence the particle occupations on different sites are independent.
For the fermions the occupation of a single site is given by a \textit{Bernoulli distribution} \cite{Derrida98_ASEP_Review_SI,Kriecherbauer10pedestrian_SI} with
\begin{equation} \label{eq:p0_ferm}
	p_{0,-}(n_i) =
	\begin{cases}
		\frac{N}{L} & n_i = 1 \\
		1-\frac{N}{L} & n_i = 0 .
	\end{cases}
\end{equation}
In the case of bosons and the ZRP the distribution is more complicated
\begin{equation} \label{eq:p0_bose}
	p_{0,+}(n_i = j) 
	= 
	\frac{1}{Z_{+}} \binom{L+(N-j)-2}{N-j}
	\simeq 
	\frac{1}{1+\bar{n}} e^{-\beta j},
\end{equation}
where $0\le j \le N$.
In the limit of $L \gg 1$ and $j \ll L$, can be well approximated by a \textit{Gibbs} measure with inverse temperature $	\beta = \log(1+\bar{n}) - \log(\bar{n})$ where $\bar{n} = N/L$.\\

From this the mean occupation as well as the fluctuations are readily obtained
\begin{equation} \label{eq:p0_bose}
	\langle n_i \rangle_0 = \bar{n}
	\qquad
	\text{and}
	\qquad
	\Delta n_i^2 = \langle n_i^2 \rangle_0 - \langle n_i \rangle_0^2
	= \bar{n}(1+\sigma \bar{n}) + O(L^{-1}),
\end{equation}
where $\langle \ldots \rangle_0$ denotes the steady state expectation value.
Two quantities that will be of interest for the following are the 
connected correlation function of the occupation
\begin{equation}
	\langle n_{i+x} n_i \rangle_0 - \langle n_{i+x}\rangle_0 \langle  n_i \rangle_0
	=
	\bar{n}(1+\sigma \bar{n}) \delta_{x,0},
\end{equation}
which turns out to be short-range correlated, as well as the current
\begin{equation} \label{eq:SI_J}
	\langle J_{i+\frac{1}{2}} \rangle_{0} 
	= 
	\gamma\langle n_i(1+\sigma n_{i+1})\rangle_{0} 
	= 
	\gamma \bar{n}(1+\sigma \bar{n}) + O(L^{-1}).
\end{equation}
From this we obtain the mean and the variance of the hydrodynamics variable, Eq.~\eqref{eq:hydro_avg}, in the steady state $n(x) = n(t\rightarrow \infty, x)$, which yields
\begin{equation} \label{eq:Asigma}
	\langle n(x) \rangle = \bar{n}
	\qquad
	\text{and}
	\qquad
	\langle n^2(x) \rangle - \langle n(x) \rangle^2
	= \frac{\bar{n}(1+\sigma \bar{n})}{\ell} = \frac{A_\sigma}{\ell}.
\end{equation}
Since the steady states for bosons and fermions are product measures, the probability distribtion for $n(x)$ is obtained from the central limit theorem and for $\ell \gg 1$ we obtain
\begin{equation} \label{eq:P0_micro}
	P_0[n(x)] 
	= 
	B \exp\left( -\sum_{x} \ell \frac{(n(x) - \bar{n})^2}{2 A_{\sigma}}\right)
	\longrightarrow
	B \exp\left( -\frac{1}{2 A_\sigma}\int\mathrm{d}x \,(n(x) - \bar{n})^2\right).
\end{equation}
The expression after the arrow corresponds to the continuum notation of the steady state measure upon identifying $\ell = \mathrm{d}x$.

\subsubsection{Current Density Relation, Euler Equation and Noise}
The current in Eq.~\eqref{eq:SI_J} suggests that for $L\gg 1$, current and density are related by
\begin{equation} \label{eq:Current_Density_rel}
	J = J(n) = \gamma n(1+\sigma n),
\end{equation}
which is known as the steady state \textit{current-density relation} if $n = \bar{n}$.
Let us consider deformations of the steady state of the form
\begin{equation}
	n(t,x) = \bar{n} + \delta n(t,x),
\end{equation}
where $\delta n(t,x)$ is not necessarily small but only varies slowly on scales $\ell \gg 1$.
In this case we expect that locally the configuration is steady state like and that the current-density relation is still valid.
Note that $n(t,x)$ is the hydrodynamic variable defined in Eq.~\eqref{eq:hydro_avg} and its dynamics are given by an \textit{Euler}-equation \cite{Kipnis98_book_SI}
\begin{equation} \label{eq:Euler}
	\partial_t n(t,x) + \partial_x J(n(t,x)) = 0.
\end{equation}
In our case this yields the (noiseless) \textit{Burgers equation}
\begin{equation} \label{eq:BE}
	\partial_t  n(t,x) + \gamma[1 + 2\gamma\sigma  n(t,x)] \partial_x  n(t,x) = 0,
\end{equation}
given that $n(t,x)$ varies slowly around the steady state. 
We note that this nonlinear equation, while capturing the dynamics on the hydrodynamic scales, is deterministic.
We model the interaction of our hydrodynamics field $n(t,x)$ with non-conserved quantities of any kind by penomenologically including some noise $\xi(t,x)$, which must be supplemented by dissipation. 
This yields the stochastic current density relation
\begin{equation} \label{eq:SI_J_noisy}
	J(n) = \gamma n(1+\sigma n) - \nu_\sigma \partial_x n - \xi.
\end{equation}
The noise arises from interaction with non-conserved quantities, which are correlated on lengths much smaller than our hydrodynamics length scales $\ell$ and  times $\tau$.
On these scales these interactions are nearly independent and therefore we model these effects in terms of a Gaussian white space-time noise characterised by
\begin{equation} 
	\langle \xi(t,x)\rangle = 0 
	\qquad
	\text{and}
	\qquad
	\langle \xi(t,x) \xi(t^\prime, x^\prime) \rangle = D_\sigma\delta(t-t^\prime) \delta(x-x^\prime),
\end{equation}
with diffusion constant $\nu_\sigma$ and noise correlation $D_\sigma$.
Plugging the current in Eq.~\eqref{eq:SI_J_noisy} into Eq.~\eqref{eq:Euler} we obtain the \textit{stochastic Burgers equation}
\begin{equation} \label{eq:SBE_app}
	\partial_t  n(t,x) + \gamma [1+2\sigma n(t,x)]\partial_x n(t,x) = \nu \partial_x^2 n(t,x) -\partial_x \xi(t,x),
\end{equation}
identical to Eq.~\eqref{eq:SBE} from the main text.
We should however keep in mind that $n(t,x)$ are mere deforemations around the steady state and the equation of motion for these are given by
\begin{equation}\label{eq:SBE_delta_n}
	\partial_t \delta n(t,x) + \gamma(1+2\sigma \bar{n}) \partial_x \delta n(t,x) + 2\gamma\sigma \delta n(t,x) \partial_x \delta n(t,x) = \nu \partial_x^2 \delta n(t,x) -\partial_x \xi(t,x).
\end{equation}

\subsubsection{Relaxation to Steady State}
In order to fix the constants $D_\sigma$ and $\nu_\sigma$ we demand that they let us thermalize to the correct steady state from Eq.~\eqref{eq:P0}. 
In addition to the exact knowledge of the steady state, Eq.~\eqref{eq:SBE_delta_n} can also be solved exactly in one dimension.
Although this a textbook exercise we show it here to be self contained.
We start by reviewing that every stochastic differential equation
\begin{equation}
	\partial_t  n(t,x)  = F[ n(t,x)] + \zeta(t,x)
\end{equation}
can be expressed in terms of a (functional) Fokker-Planck equation.
Here we replaced $\delta n \rightarrow n$ for notational convenience.
We also defined the \textit{conserved noise} $\zeta(t,x) = \partial_x \xi(t,x)$ with correlations
\begin{equation}
	\langle \zeta(t,x)\zeta(t^\prime,x^\prime) \rangle
	=
	\langle \partial_x \xi(t,x) \partial_{x^\prime} \xi(t^\prime,x^\prime)\rangle
	=
	-D_\sigma\delta(t-t^\prime) \partial_x^2 \delta(x-x^\prime).
\end{equation}
The Fokker-Planck equation reads 
\begin{eqnarray} \label{eq:FPE}
	\partial_t P_t[ n] 
	& = & 
	\int \mathrm{d}x \,
	\frac{\delta}{\delta n(x)} 
	\left\{
	- F_1[n] P_t[n]
	- 
	F_0[n]P_t[n] 
	+ 
	\frac{\delta}{\delta n(x)} \left( -\frac{D_\sigma}{2} \partial_x^2 P_t[n]\right)
	\right\},
\end{eqnarray}
where we defined $F_0[ n] = \nu_\sigma \partial_x^2 n$ and $F_1[ n] = -(\gamma(1+2\sigma \bar{n}) + 2\sigma \gamma n) \partial_x  n $. 
We will now show that a steady state, much like the one in Eq.~\eqref{eq:P0_micro}, of the form
\begin{equation} \label{eq:P0_SBE}
	P_0[n] = C e^{-\frac{\nu_\sigma}{D_\sigma}\int\mathrm{d}x^\prime n^2(x^\prime)},
\end{equation}
is a steady state solution. 
We start by computing the variational derivative $	\frac{\delta}{\delta n(x)}  P_0[n] = -\frac{2\nu_\sigma}{D_\sigma} n(x) P_0[n]$, which shows that the second and the third term in Eq.~\eqref{eq:FPE} cancel.
This leaves us with showing that the integral over the first term in Eq.~\eqref{eq:FPE} vanishes. 
The variational derivative yields an expression of the type
\begin{equation}\label{eq:tot_deriv}
	\begin{split}
		\frac{\delta }{\delta n(x)}F_1[n]P_0[n] & = P_0[n](a\partial_x n(x) + b n(x)\partial_x n(x) + c n^2(x)\partial_x n(x)) \\
		& =
		P_0[n]
		\left( 
		a\partial_x n(x) + b \partial_x \left( \frac{1}{2}n^2(x) \right) + c \partial_x \left( \frac{1}{3} n^3(x)\right)
		\right),
	\end{split}
\end{equation}
which is expressed in terms of a total derivative.
For the conserved quantity for a ring $n(t,x) = n(t,x+L)$, much like for the microscopic solution in Sec.~\ref{sec:SS}, the integral over the expression in Eq.~\eqref{eq:tot_deriv} vanishes, which shows that Eq.~\eqref{eq:P0_SBE} is indeed a steady state.
Going back to displaced variables $n \rightarrow \bar{n} + \delta n$ we obtain the final expression of the steady state
\begin{equation} \label{eq:P0_SBE_final}
	P_0[n] = C e^{-\frac{\nu_\sigma}{D_\sigma}\int\mathrm{d}x^\prime (n(x^\prime) - \bar{n})^2}.
\end{equation}
Comparing this with Eq.~\eqref{eq:P0_micro} we obtain the identity
\begin{equation} \label{eq:fluct_diss_NLFHD}
	D_\sigma
	=
	2\nu_\sigma \bar{n}(1+\sigma \bar{n}).
\end{equation}
This shows that there is still one free parameter left after fixing either $D_\sigma$ or $\nu_\sigma$.
The analytical predicitions in the main part of the paper do not depend on the specifics of this choice.

\subsection{SBE from a Fokker-Planck Approximation}
\label{SI:FPE_approx}

In this section we outline another method to motivate the emergence of the stochastic Burgers equation~\eqref{eq:SBE} using a Fokker-Planck approximation starting from Eq.~\eqref{eq:ME_class} of the main text. 
This can only be trusted in the limit of bosonic particles and the ZRP, at high occupation numbers $1 \ll n_i = n_0 \tilde{n}_i$ where $n_0 \gg 1$ and $\tilde{n}_i \sim 1$.
Within the Fokker-Planck approximation we neglect ``finer" details of the dynamics of the local distribution $P(t,\ldots,n_i,\ldots)$, which are on scales much smaller than $n_0$.
The shift operators of the Eq.~\eqref{eq:ME_class} are expanded up to second order and we obtain
\begin{equation} \label{eq:FP_approx}
	T_i^{\pm} 
	= 
	e^{\pm \partial_{n_i}} 
	=
	e^{\pm n_0^{-1}\partial_{\tilde{n}_i}} 
	=
	1 
	\pm \partial_{n_i} 
	+ \frac{1}{2}\partial_{n_i}^2
    + O(n_0^{-3}).
\end{equation}
The contributions we neglect will give corrections on the order $n_0^{-3}$.
This truncation is referred to as the \textit{Fokker-Planck approximation} since it generically yields a diffusive stochastic process as we show in the following.

Inserting Eq.~\eqref{eq:FP_approx} up to second order in Eq.~\eqref{eq:ME_class}, we obtain the Fokker-Planck operator 
\begin{equation} \label{eq:L1L2}
\dot{P}(t,\vec{n}) = (L_1 + L_2)P(t,\vec{n}).
\end{equation}
On the side we mention that in the main text we briefly commented that as long as $\gamma_+ \ne \gamma_-$ the difference to the totally asymmetric case will be only qualitative. 
Here we show this explicitly by keeping both hopping rates which we decompose as $\gamma_\pm = \nu_\sigma + \frac{v}{2}$ where $\sigma = 0,+$ in the following.

The first part of the Fokker-Planck operator in Eq.~\eqref{eq:L1L2} has the form
\begin{equation}
	\begin{split} \label{eq:L1}
		L_1 
		& =  
		\sum_{i=1}^N 
		\left(\nu_\sigma + \frac{v}{2}\right) [\partial_{i} - \partial_{{i+1}}] n_{i}(1+ \sigma n_{i+1}) + \left(\nu_\sigma-\frac{v}{2}\right) [\partial_{i} - \partial_{{i-1}}] n_{i}(1+\sigma n_{i-1})  \\
		& =
		-\sum_{i=1}^L \partial_{i}
		\left( 
		-v \left( 1 +2 \sigma n_i \right)\frac{n_{i+1}-n_{i-1}}{2} + \nu_\sigma (n_{i+1} - 2n_i + n_{i-1})
		\right),
	\end{split}
\end{equation}
where we used $\partial_i \equiv \partial_{n_i}$ for a simpler notation.
The totally asymmetric case with $\gamma_- =0$ corresponds to $2\nu_\sigma = v =\gamma$.
We proceed by introducing the discrete derivative/gradient operators
\begin{equation}
	(\nabla)_{ij} 
	=
	\frac{1}{2} \left( 
	-\delta_{i,j+1} +\delta_{i,j-1} - \delta_{i,j-N+1} -  \delta_{i-N+1,j} 
	\right)
\end{equation}
and the discrete Laplacian matrix
\begin{equation}
	(\Delta)_{ij} = -2 \delta_{i,j} +\delta_{i,j+1} +\delta_{i,j-1} + \delta_{i,j-N+1} +  \delta_{i-N+1,j}.
\end{equation}
Note that here we have assumed periodic boundary conditions, which, however, is not crucial.
In this notation the Eq.~\eqref{eq:L1} is now cast to a more compact form
\begin{equation}
	L_1 = -\sum_{i=1}^L \partial_{i}
	\underbrace{ 
		\left(
		-v \left( 1 + 2 \sigma n_i \right)(\nabla \cdot \vec{n})_i + \nu_\sigma (\Delta \cdot \vec{n})_i
		\right)
	}_{=(A[\vec{n}])_i}.
\end{equation}
The second term in the expansion Eq.~\eqref{eq:L1L2} is more lengthy and therefore we introduce the shorthand notation
\begin{eqnarray}
	(J^{+})_{ij} & = & \frac{\nu_\sigma}{2}(J_{i,i+1} + J_{i+1,i}) \delta_{i,j} \\
	(J^{-})_{ij} & = & \frac{v}{4}(J_{i,i+1} - J_{i+1,i}) \delta_{i,j} \\
	J_{a,b} & = &  n_a(1+ \sigma n_b).
\end{eqnarray}
In this notation the second term in the Fokker-Planck operator reads
\begin{equation}
	\begin{split}
		L_2 = 
		&
		\quad
		\frac{1}{2}\sum_{i=1}^N \nu_\sigma \left[\partial_{i}^2 \left( J_{i,i+1} + J_{i+1,i} + J_{i,i-1} + J_{i-1,i} \right) - 2 \partial_{{i}}\partial_{{i+1}}\left(J_{i,i+1} + J_{i+1,i}\right)\right] \\
		&+
		\frac{1}{2}\sum_{i=1}^N
		\frac{v}{2}\left[\partial_{i}^2 
		\left(
		J_{i,i+1} - J_{i+1,i} + J_{i,i-1} - J_{i-1,i}
		\right) 
		-2 \partial_{i}\partial_{{i+1}}\left(J_{i,i+1} - J_{i+1,i}\right)\right] \\
		= & 
		\frac{1}{2}\sum_{i,j}
		\partial_{i}\partial_{j}
		\underbrace{\left[2 \nabla_+ \cdot (J^+ + J^-)\cdot \nabla_+^\top \right]_{ij}}_{=(B[\vec{n}]\cdot B^\top[\vec{n}])_{ij}},
	\end{split}
\end{equation}
where obtained the last equality by introducing the discrete forward derivate operator
\begin{equation}
	(\nabla_+)_{ij} = -\delta_{i,j} + \delta_{i,j-1} + \delta_{i-N+1,j}.
\end{equation}
Then the Fokker-Planck equation attains a more standard form, 
\begin{equation}
	\dot P(t,\vec{n}) = -\sum_{i}\partial_{n_i}\left((A[\vec{n}])_i P(t,\vec{n})\right) 
	+ 
	\frac{1}{2}\sum_{ij} \partial_{n_i}\partial_{n_j} (B[\vec{n}]\cdot B^\top[\vec{n}])_{ij} P(t,\vec{n}),
\end{equation}
where we have undone $\partial_i \rightarrow \partial_{n_i}$. The corresponding stochastic differential equation is
\begin{equation} \label{eq:SBE_discrete}
	\dot{n}_i(t) 
	= 
	(A[\vec{n}(t)])_i + \zeta_i(t)  
	= 
	-v \left( 1 + 2 \sigma n_i \right)(\nabla \cdot \vec{n})_i + \nu_\sigma (\Delta \cdot \vec{n})_i
	+
	\zeta_i(t).
\end{equation}
The noise correlation are of the form
\begin{equation}\label{eq:Noise_1}
	\langle \zeta_i(t) \zeta_j(t^\prime)\rangle 
	= 
	[2\nabla_+\cdot(J^+ + J^-)\cdot \nabla_+^\top]_{ij}\delta(t-t^\prime).
\end{equation}
Note the correlations in time are white (due to the Fokker-Planck approximation), but in real space they crucially depend on the density $n_i$.

In the case of the steady state, \eqref{eq:P0_micro}, or also for small deformations around a state that is locally in equilibrium, we have
$J_{i,i+1}\simeq J_{i+1,i}\simeq J_{i,i}$ which implies $(J^-)_{ij} \simeq 0$ and $(J^+)_{ij} \simeq \nu_\sigma n_i(1+n_i)\delta_{i,j}$.
In this limit and in regions where the average density is sufficiently smooth, we can neglect small density variations in the second part of the Fokker-Planck operator and obtain
\begin{equation}
	L_2 \simeq \sum_{i,j} \partial_{n_i}\partial_{n_j}(-2\nu_+ n_i(1+\sigma n_i)\Delta_{ij}).
\end{equation}
Here we used the identity $\nabla_+\cdot \nabla_+^\top = -\Delta$.
This implies that the noise correlations in Eq.~\eqref{eq:Noise_1} become
\begin{equation}
	\langle \zeta_i(t) \zeta_j(t^\prime)\rangle  
	=
	-2\nu_\sigma n_i (1+\sigma n_i)\Delta_{i,j}\delta(t-t^\prime).
\end{equation}
Equivalently, the noise can be represented in terms of the derivatives of a white space-time noise
\begin{equation}
	\vec{\zeta}(t) = \nabla_+ \cdot \vec{\xi}(t) 
	\qquad
 \text{with}\qquad
	\langle \xi_i(t)\xi_j(t^\prime)\rangle = \delta_{ij}\delta(t-t^\prime).
\end{equation}
Taking the naive continuum limit, which means identifying $n_i(t) \rightarrow n(t,x)$, $\nabla \rightarrow \partial_x$ and $\Delta \rightarrow \partial_x^2$, we obtain from Eq.~\eqref{eq:SBE_discrete} the stochastic Burgers equation
\begin{equation}\label{eq:SBE_FPA}
	\partial_t n(t,x) + v(1+2\sigma n(t,x))\partial_x n(t,x) = \nu_\sigma \partial_x^2 n(t,x) + \partial_x \xi(t,x)
\end{equation}
with white space-time noise
\begin{equation}
	\langle \xi(t,x)\xi(t^\prime,x^\prime)\rangle 
	= 
	\underbrace{\gamma n(x,t)(1+\sigma n(x,t))}_{ = D_\sigma(x)} \delta(x-x^\prime)\delta(t-t^\prime).
\end{equation}
For a local density of $n(x,t)\simeq \bar n$ this derivation yields the following identity connecting fluctuations and dissipation 
\begin{equation}
 D_\sigma = \gamma \bar{n}(1+\sigma \bar{n}) = 2\nu_\sigma 
\bar{n}(1+\sigma \bar{n}),
\end{equation}
which is the same as the one obtained in Eq.~\eqref{eq:fluct_diss_NLFHD}. However, the derivation in terms of the Fokker-Planck equation also fixes the value of $\nu_\sigma=\gamma/2$.

We emphasize that the derivation of Eq.~\eqref{eq:SBE_FPA} presented in this subsection is based on two crucial approximations: (i) The assumption of smooth particle distribution in each lattice site with a large average particle number, and (ii) the assumption of a slowly varying density profile, which allows us to perform the continuum limit $n_i(t)\rightarrow n(x,t)$. Despite an obvious violation of assumption (i) in the examples studied in the main text, we find that the predictions of Eq.~\eqref{eq:SBE_FPA} are fully consistent with the exact steady states discussed above for both fermions and bosons and agree as well as with numerical simulations of Eq.~\eqref{eq:ni}. Therefore, we can argue that Eq.~\eqref{eq:SBE_FPA} is more generally valid to describe both fermionic and bosonic transport processes in situations where the system develops a density profile that is smooth and close to a local equilibrium. 
Note that this situation can be violated, for example, in the case of bosons in the presence of boundaries, where rapidly alternating density profiles can appear \cite{Garbe23_SI}.

\subsection{Mean Number of Transported Particles}
\label{SI:Nt}

In this section we are going to compute the first moments of $\langle \mathcal{N}(t)\rangle$
from the noiseless Burgers equation
\begin{equation}
	\partial_t n(t,x) + \gamma(1+2\sigma n(t,x))\partial_x n(t,x) = 0,
\end{equation}
starting from the initial condition
\begin{equation}
	n(t=0,x) = 
	\begin{cases}
		1 & x \le 0 \\
		0 & x > 0.
	\end{cases}
\end{equation}
This must be seen as a smoothened version of the actual step initial condition that we consider on the level of particles. 
The solution of the initial value problem is obtained using the method of characteristics. 
We restate here the current-density relation for later purposes
\begin{equation}
	J(n) = \gamma n(1+\sigma n).
	\label{eq:currdenbis}
\end{equation}

For fermions, $\sigma = -1$, we obtain a \textit{rarefaction wave} \cite{Krapivsky10_book_kinetic_SI,Kriecherbauer10pedestrian_SI}
\begin{equation}
	n(t,x) 
	=
	\begin{cases}
		1 & x < -\gamma t \\
		\frac{1}{2}\left(1-\frac{x}{vt}\right) & \vert x\vert \le \gamma t \\
		0 & x > \gamma t
	\end{cases}
\end{equation}
with a constant density in the origin $n(t,x=0) = \frac{1}{2}$, which gives the current $J(n(t,x=0)) = \gamma/4$.\\

For bosons, $\sigma = +1$, the characteristics intersect and we obtain a \textit{shock wave}:
\begin{equation}
	n(t,x) =
	\begin{cases}
		1 & x \le  c_st \\
		0 & x > c_st
	\end{cases}
	\qquad
	\text{with}
	\qquad
	c_s =2 \gamma,
\end{equation}
The shock speed was obtained from mass conservation or more formally the Rankine-Hugoniot criterion:
\begin{equation}
	c_s = \frac{J(n_-) - J(n_+)}{n_- - n_+} = \gamma (1 + n_-) = 2\gamma
\end{equation}
where $n_+ = 0$ and $n_- = 1$, which correspond to the height of the hydrodynamic profiles to the left and right of the shock front, respectively. At the origin, the height is $n(t,x=0)= 1$ at all times, and the current is therefore $J(n(t,x=0)) = 2\gamma$.
\\

For the ZRP, $\sigma = 0$, the characteristics are all parallel and we still obtain a step profile moving to the right:
\begin{equation}
	n(t,x) =
	\begin{cases}
		1 & x \le  \gamma t \\
		0 & x > \gamma t,
	\end{cases}
\end{equation}
We have $n(t,x = 0) = 1$ in this case as well, and $J(n(t,x=0)) = \gamma$.\\

By integrating in time the current at the origin $J(n(t,x=0))$, we obtain the total number of particles jumping through the origin, which in turns gives us the total population on the right side of the chain. We obtain

\begin{equation}
	\langle \mathcal{N}(t)\rangle = \int_0^t\mathrm{d}\tau\,J(n(\tau,x=0))
	= j_{0,\sigma}t,
	\qquad
	\text{with}
	\qquad
	j_{0,\sigma}
	=
	\begin{cases}
		\frac{\gamma }{4} & \sigma = -1 \\
		\gamma & \sigma = \;\;\,0 \\
		2\gamma & \sigma = +1.
	\end{cases}
\end{equation}
This result gives us the overall linear increase depicted by the dashed lines on Fig.~\ref{fig:2}(a) of the main paper.

\section{Supplement to paragraph ``KPZ universality"}
\label{SI:KPZ_univ}

In this paragraph we review the KPZ scaling exponents in Sec.~\ref{SI:Scaling_KPZ}, which are known exactly in one dimension. 
Then in Sec.~\ref{SI:Family_Vicsek} we tie this to particle transport which provided the analytical curves in Fig.~\ref{fig:2}(b-c).

\subsection{Scaling of Edwards-Wilkinson and Kardar-Parisi-Zhang Equation} \label{SI:Scaling_KPZ}

In this section we give a derivation of the scaling exponents characterizing the transport fluctuations.
We start by considering Eq.~\eqref{eq:KPZ_transp} of the main text and transforming away the advection term $-\gamma \partial_x h(t,x)$ by a Galilei transformation $x \rightarrow x^\prime = x- \gamma t$ and $t\rightarrow t^\prime = t$.
This yields the KPZ equation for $\tilde{h}(t,x) = h(t,x-\gamma t)$,
\begin{equation} \label{eq:KPZ_no_adv}
	\partial_t h(t,x) = \sigma \gamma (\partial_x h(t,x))^2 + \nu \partial_x^2 h(t,x) + \xi(t),
\end{equation}
where we have omitted the tilde $\tilde{h} \rightarrow h$.
The noise term transforms to $\tilde{\xi}(t,x) = \xi(t,x+\gamma t)$ but since the noise fluctuations are unaffected
$\langle \tilde{\xi}(t_1,x_1) \tilde{\xi}(t_2,x_2)\rangle = D \delta(t_1-t_2)\delta(x_1-x_2)$ we set $\tilde{\xi} \rightarrow \xi$ for a simpler notation.

Under the premise that the KPZ equation is a critical theory and describes the physics at all, especially the long scales it should be invariant under a rescaling of space and time
\begin{equation} \label{eq:rescaling}
	t \rightarrow b^z t
	\qquad
	x \rightarrow b x,
\end{equation}
with $b>1$.
We remark that this will affect the advection term which must be removed by a different Galilei transformation depending on $b$. 
The rescalings of space and time will carry over to the field which transforms to
\begin{equation} \label{eq:balpha_h}
	h(t,x) \rightarrow h(b^z t, bx) = b^\alpha h(t,x).
\end{equation}
Applying this rescaling to the KPZ equation~\eqref{eq:KPZ_no_adv} we obtain
\begin{equation} \label{SI_eq:KPZ_resc}
	\partial_t h(t,x) 
	= 
	+
	b^{\alpha + z -2}\gamma \sigma  (\partial_x h(t,x))^2 
	+
	b^{z-2} \nu \partial_x^2 h(t,x)
	+ b^{\frac{z-1}{2}-\alpha}\xi(t,x).
\end{equation}
The simple rescaling analysis we performed this yields is identical to a more sophisticated renormalization group (RG) treatment (up to \textit{tree-level}) without including corrections from the nonlinearity (\textit{1-loop} and higher). 
For the EW equation this treatment is complete while due to the nonlinearity of the KPZ, corrections need to be considered which alter the results from simple rescaling.

\subsubsection{Scaling Exponents of the EW Equation}

In the case of the Edward-Wilkinson equation $\sigma = 0$, the theory on coarse grained space and time scales will have a renormalized diffusion and fluctation constant
\begin{equation} \label{eq:renorm_alpha_z}
	\nu \longrightarrow b^{z-2}\nu 
	\qquad
	D \rightarrow b^{z-1-2\alpha}D.
\end{equation}
This gives two equations for the two unknown scaling exponents and solving this set of equations we obtain
\begin{equation}
	\alpha_{\mathrm{EW}} = \frac{1}{2}
	\qquad
	z_{\mathrm{EW}} = 2.
\end{equation}
The diffusion and fluctuation constant will not change (\textit{renormalize}) under coarse graining.

\subsubsection{Scaling Exponents of the KPZ Equation}
We now turn to the case of $\sigma = \pm 1$ where the scaling exponents can, a priori not be obtained by the same simplified analysis outlined above.
Yet under the simple rescaling the diffusion and fluctuation constant will change (at tree level) as shown in Eq.~\eqref{eq:renorm_alpha_z} and the KPZ nonlinearity as
\begin{equation}
\sigma \gamma \rightarrow \sigma \gamma b^{\alpha+z-2}.
\end{equation}
Rescaling will yields three equations for two exponents.
In case of the KPZ equation in one dimension there is a alternative road to finding the critical exponents besides perturbative RG \cite{Forster77_SI,book_Kamenev2023field_SI}, which we will quickly outline following \cite{Barabasi95boook_SI,Kardar07book_SI} with a few more details.

The first key observation to obtain the scaling exponents without RG, is that the KPZ equation has an invariance under \textit{tilting}.
More concretely given the solution $h(t,x)$ another solution is constructed based on it by
\begin{equation}
	h^\prime(t,x) = h(t,x-ut) - \frac{u}{2\sigma\gamma}  x+ \frac{u^2}{4\sigma\gamma}t,
\end{equation}
which is checked by plugging it in the KPZ equation.
Note that this solution has the initial condition $h^\prime(t=0,x) = h(0,x) - \frac{u}{2\sigma \gamma}x$ which is tilted with respect to the original initial condition.
Importantly we notice that this invariance under tilts explicitly contains $\gamma$ and if we demand that a tilt transformation is well defined under/independent with respect to rescaling $b$, we have to demand that $\gamma$ does not change under a rescaling. 
This must be true at all levels of the perturbative renormalization, but especially at tree level, accessible by simple scaling analysis and therefore 
\begin{equation}\label{eq:Ward}
\alpha + z  =2.
\end{equation}
Using more sophisticated techniques it was shown that this symmetry saveguards this identity in all dimensions and is non-perturbative.
So in case we found $\alpha$ this would also fix $z$.

In order to proceed we will have to briefly discuss the role of $\alpha$ on real space correlations functions of the form
\begin{equation} 
	G(x) = \langle (h(x) - h(0))^2 \rangle_0 -  \langle (h(x) - h(0)) \rangle_0^2
\end{equation}
where $\langle \ldots \rangle_0$ corresponds to steady state average.
Rescaling space and time yields the functional equation
\begin{equation}
	G(bx) = b^{2\alpha}G(x)
\end{equation}
which is solved by
\begin{equation} \label{eq:Gx_alpha}
G(x) = C \vert x \vert^{2\alpha}.
\end{equation}

In $D=1$ the KPZ equation can be rewritten in terms of a Fokker-Planck equation and solved with the exact same arguments as the stochastic Burgers equation if periodic boundary conditions are imposed.
Since we have already done the work above one may just set $\delta n = \partial_x h$ in Eq.~\eqref{eq:P0_SBE} and obtain
\begin{equation} \label{eq:P0_KPZ}
	P_0[h] = C e^{-\frac{\nu_\sigma}{D_\sigma}\int\mathrm{d}x^\prime (\partial_x h(x))^2}
 \implies
 G(x) = \frac{D_\sigma}{2\nu_\sigma} \vert x \vert
\end{equation}
where $h(x) = h(t\rightarrow \infty,x)$ from which we read off $\alpha =\frac{1}{2}$.

Alternatively and to show consistency with the underlying microscopic particle picture especially for the unexplored case of the bosons the same can be shown starting from the exactly known steady state on the ring.
We consider the difference in the site to site transport
\begin{equation}
h_{i+\frac{1}{2}} - h_{i-\frac{1}{2}} = n_i.
\end{equation}
Adding a constant of $ \frac{1}{2}$ as we later do for the corner initial conditions will not change the argument.
According to Eq.~\eqref{eq:p0_bose} and Eq.~\eqref{eq:p0_ferm} the local distribution $n_i$ are independent and identically distributed with
$\langle n_i \rangle = \bar{n}$ and $\langle n_i^2 \rangle - \langle n_i \rangle^2 = \bar{n}(1+\sigma \bar{n})$.
Let us now consider the difference
\begin{align}
h_{x + \frac{1}{2}} - h_{0+\frac{1}{2}} & = \underbrace{h_{x+\frac{1}{2}} - h_{x-1 + \frac{1}{2}}}_{=n_x}
 + \underbrace{ h_{x-1 + \frac{1}{2}} + h_{x-2 + \frac{1}{2}}}_{=n_{x-1}} 
 +\ldots  + 
 \underbrace{h_{1 + \frac{1}{2}} - h_{0+\frac{1}{2}}}_{=n_1} 
  = \sum_{j=1}^x n_x.
\end{align}
We now apply the central limit theorem for $x \gg 1$ we obtain the distribution
\begin{equation}
P(h_{x + \frac{1}{2}} - h_{0+\frac{1}{2}}) = \frac{1}{\sqrt{2\pi x \sigma_\delta^2}} \exp\left( -\frac{( h_{x + \frac{1}{2}} - h_{0+\frac{1}{2}} - x \bar{n})^2}{2 \vert x\vert(\bar{n}(1+\sigma \bar{n})))^2} \right).
\end{equation}
An hence we obtain the same result as for the KPZ equation
\begin{equation}
G(x) 
= 
\langle (h_{x + \frac{1}{2}} - h_{0+\frac{1}{2}})^2 \rangle 
- 
\langle h_{x + \frac{1}{2}} - h_{0+\frac{1}{2}}\rangle^2
=
\vert x\vert  (\bar{n}(1+\sigma \bar{n}))^2.
\end{equation}
Together with the identity in Eq.~\eqref{eq:Ward} we obtain
\begin{equation}
\alpha_{\mathrm{KPZ}} = \frac{1}{2}
\qquad
\implies
\qquad
z_{\mathrm{KPZ}} = \frac{3}{2}.
\end{equation}

\subsection{Family-Vicsek Scaling} \label{SI:Family_Vicsek}

In this section we discuss the scaling of the number of particles transported $\mathcal{N}(t)$.
In the main text we made the connection of the transport variables and in the hydrodynamic limit we obtain $\mathcal{N}(t) = h(t,x=0)$.
The tansport fluctuations are defined as
\begin{equation}
	\Delta \mathcal{N}(t) = \sqrt{\langle h^2(t,0)\rangle - \langle h(t,0)\rangle^2}.
\end{equation}
Rescaling this equation we obtain the functional equation
\begin{equation}
	\Delta \mathcal{N} (b^z t)= b^{\alpha}\Delta \mathcal{N} (t)
\end{equation}
which can be solved up to a constant and which gives \cite{Barabasi95boook_SI,Taeuber_14_SI} 
\begin{equation}
	\Delta \mathcal{N}(t) = C t^{\alpha/z} = C t^\beta.
\end{equation}
From this scaling analysis we obtain
\begin{equation}
    \beta_{\mathrm{EW}} = \frac{1}{4}
    \qquad
    \beta_{\mathrm{KPZ}} = \frac{1}{3}.
\end{equation}
The numerical results for bosons and fermions are compared with these theoretical predictions in Fig.~\ref{fig:2}(b) of the main text.

The fluctuations of the KPZ equation are determined by three scaling exponents but focusing on transport we only resolve the ratio $\alpha/z$.
In the following we consider a quantity \cite{Family85_SI}, which remedies this and which is readily compared with numerics.
We start by considering and overall transport (interface height) average of over all (dual) sites in the system
\begin{equation}
	w(t,L) = \sqrt{\frac{1}{L}\sum_{i=1}^L \langle(h(t,x_i) - \bar{h}(t))^2\rangle}\qquad \bar{h}(t) = \frac{1}{L} \sum_{i=1}^L h(t,x_i).
\end{equation}
Rescaling of space $L\rightarrow b L$ and time yields the functional equation
\begin{equation}
	w(b^z t , b L) = b^\alpha w(t,L).
\end{equation}
The formal solution of this is
\begin{equation} 
	\label{eq:scalingform}
	w(t,L) = L^\alpha f\left( \frac{t}{L^z}\right).
\end{equation}
While we cannot uniquely find the solution, we can fix it up to an unknown function $f(x)$.
Invoking some simple physical arguments we will now infer the asymptotics of $f(x)$ and consequently $w(t,L)$.

We start identifying the timescale $t_{\times} \equiv L^z$ (leaving out a constant which corrects the dimensions).
For times $t \ll t_\times$ we expect that $w(t,L)\sim t^\beta$ similar to our result $\Delta \mathcal{N}(t) \sim t^\beta$ and be system size independent.
This is satisfied when choosing $f(x\ll 1) \sim x^\beta$.
At times $t \gg t_\times$ the real space correlations are upper bounded by system size $G(L) = L^{2\alpha}$ [see Eq.~\eqref{eq:Gx_alpha}] and therefore expect that $w(t\gg t_\times,L) \sim L^\alpha$ saturates at large times.
In summary these considerations fix the asymptotic behavior of the unknown functions  \cite{Family85_SI} and we obtain
\begin{equation}
	f(x) = 
	\begin{cases}
		x^\beta & x \ll 1 \\
		\mathrm{const.} & x \gg 1
	\end{cases}
\end{equation}
from which we obtain
\begin{equation}
	w(t,L) \sim
	\begin{cases}
		t^\beta & t \ll L^z \\
		L^\alpha & t \gg L^z
	\end{cases}.
\end{equation}
Note that some solvable microscopic models in the KPZ universality class give exactly tractable predictions for $f(x)$ \cite{Praehofer02_SI} which match the simple consideration given above.
The theoretically predicted asymptotics of $w(t,L)$ are then compared with the numerical results in Fig.~\ref{fig:2}(c) and are in excellent agreement.

\section{Supplement to paragraph ``Full counting statistics"}

We start this section~\ref{SI:Gamma_dim} by briefly reviewing an argument which fixes $\Gamma_\sigma$ in Eq.~\eqref{eq:h_FCS} up to a constant.
In Sec.~\ref{SI:TWD} we go on and introduce the Tracy-Widom distribution in the context of random matrix theory.
Finally in Sec.~\ref{SI:FCS_numerics} we explain how the $\Gamma_\sigma$ was extracted numerically.

\subsection{$\Gamma$ from Dimensional Analysis} \label{SI:Gamma_dim}

In this section we constrain the $\Gamma_\sigma$ from a simple argument in the spirit of \cite{Kriecherbauer10pedestrian_SI,Takeuchi18_SI}, using dimensional analysis and the knowledge of the real space correlations in the steady state.
For the case of the Edwards-Wilkinson equation, $\sigma = 0$, we refer the reader to \cite{book_Kamenev2023field_SI}.

We start by noting that in Eq.~\eqref{eq:h_FCS} $[\eta] = 1$ and we find that $[\mathcal{N}] = [h_{\frac{1}{2}}]  = [(\Gamma t)^{1/3}]$.
Therefore our problem of finding $\Gamma$ is reduced to finding the dimensions of $[h]$ in terms of all "relevant" quantities which is as vague as it sounds.
Starting from Eq.~\eqref{eq:KPZ_transp} and focusing on the nonlinear term especially we find that
\begin{equation}
    [\gamma] = \frac{[x]^2}{[h][t]}.
\end{equation}
Another quantity which is tied to an exactly known quantity is the strength of the transport correlations in the steady state $\langle (h_{i+\frac{1}{2}} - h_{\frac{1}{2}})^2 \rangle_0 = A_\sigma \vert i \vert $ where we recall that $A_\sigma = \bar{n}(1+\sigma \bar{n})$.
From this we find that
\begin{equation}
[A_\sigma] = \frac{[h]^2}{[x]}.
\end{equation}
Assuming none of the other identities involving $[h]$ are important. 
Solving these equations for $[h]$ we obtain
\begin{equation}
h(t) = \tilde{h}(t)(C_\sigma A_\sigma^2 \gamma t)^{\frac{1}{3}}.
\end{equation}
with the dimensionless quantity $[\tilde{h}] = 1$.
Here $C_\sigma$ corresponds to a constant which is not fixed by dimensional analysis.
From this read off that
\begin{equation}
\Gamma_\sigma =  C_\sigma \gamma A_\sigma^2      
\end{equation}
up to a proportionality constant. 
In order to fix the constant $C_\sigma$ we need to either invoke exact solution as in the case of fermions $C_- = 1$ \cite{Johansson2000_SI} or have to resort on a numerical analysis (see Sec.~\ref{SI:FCS_numerics}).

\subsection{Tracy-Widom Distribution} \label{SI:TWD}

In order to be self contained we give a brief reminder on the Tracy-Widom distributions (TWD) of random matrix theory.
We consider a matrix sampled from the $H_{\mathrm{1}}$ Gaussian orthogonal ($H_{\mathrm{2}}$ Gaussian unitary ensemble) respectively where we have used the Dyson index notation $\beta = 1$ for GOE ($\beta = 2$ for GUE).
We define the GOE by considering a real $N \times N$ matrix $M$ with entries drawn from a normal distribution $M_{ij} \sim N(0,1)$ or $\langle M_{ij}M_{kl}\rangle = \delta_{ik}\delta_{jl}$ and from this we obtain
\begin{eqnarray}
	H_{1} \sim \frac{1}{2}(M + M^\top).
\end{eqnarray}
In case of the GUE we sample a complex Gaussian matrix $M_{ij} = x_{ij} + i y_{ij}$ with $x_{ij},y_{ij}\sim N(0,1)$ or equivalently with variance $\langle M_{ij}M^*_{kl}\rangle = 2\delta_{ik}\delta_{jl}$ and
\begin{eqnarray}
	H_{2} \sim \frac{1}{\sqrt{8}}(M + M^\dagger)
\end{eqnarray}
where $M_{ij}$ is complex and with vanishing mean in both cases $\langle M_{ij}\rangle = 0$.
From this we obtain the second moments
\begin{eqnarray}
\langle (H_1)_{ij} (H_1)_{kl}\rangle
=
\frac{1}{2}(\delta_{ik}\delta_{jl} + \delta_{il}\delta_{jk})
\qquad
\langle (H_2)_{ij} (H_2)_{kl}\rangle
=
\frac{1}{2}\delta_{ik}\delta_{jl} 
\end{eqnarray}
and $\langle H_\beta \rangle = 0$.

The spectrum of $H_{\beta}$ is real $\lambda_N \le \ldots \le \lambda_1$ with the largest eigenvalues $\lambda_1$.
The distribution of the largest eigenvalues $\lambda_1$ was determined in \cite{Tracy93_SI,Tracy94GUE_SI,Tracy96GOE_GSE_SI} and is expressed as
\begin{eqnarray}
\lambda_1 = \sqrt{2 N } + \frac{N^{-1/6}}{\sqrt{2}} \eta
\end{eqnarray}
where $\eta$ is sampled from the distribution $\mathrm{TWD}_{\beta}(\eta)$.

\begin{eqnarray}
\mathrm{TWD}_{\beta}(\eta) = \frac{\mathrm{d}}{\mathrm{d}\eta} F_\beta(\eta)
\end{eqnarray}
where $F_\beta$ corresponds to the cumulative distribution function $F_\beta(x) = \int_{-\infty}^x\mathrm{d}y\,\mathrm{TWD}_\beta(x)$.
The cumulative distribution function can be expressed in the compact form \cite{Ferrari_05_formula_SI}
\begin{eqnarray}
	F_\beta(x) = \mathrm{Det}(\mathbb{1}-B^\beta(x))
\end{eqnarray}
where we defined the operator $B(x) = \mathrm{Ai}(x+ x_1+x_2)$ in the variables $x_1$ and $x_2$ with the Airy function $\mathrm{Ai}(x)$.
For $\beta = 2$ we obtain $B^2(x) = \int\mathrm{d}y \mathrm{Ai}(x + x_1 + y)\mathrm{Ai}(x +  y + x_2)$ which has the form of an Airy-kernel.
The four lowest moments of the TWD are characterized by the mean and variance
\begin{equation}
	\langle \eta^n \rangle_c \equiv \langle (\eta - \langle \eta \rangle)^n \rangle
\qquad
\mathrm{Var}(\eta) = \langle \eta^2 \rangle_c =\langle \eta^2 \rangle - \langle \eta \rangle^2
\end{equation}
as well as skewness and kurtosis
\begin{equation}
	\mathrm{Sk}(\eta) \equiv \frac{\langle \eta ^3 \rangle_c}{\langle \eta^2 \rangle_c^{3/2}} 
	\qquad
	\mathrm{Ku}(\eta) \equiv \frac{\langle \eta^4 \rangle_c}{\langle \eta^2 \rangle_c^2}.
\end{equation}
The exact numerical values are given in the table below.
\begin{center}
	\begin{tabular}{|c|c|c|c|c|}
		\hline  
		$\beta$ & $\langle \eta\rangle$ & $\langle \eta^2 \rangle_c$ & $\mathrm{Sk}(\eta)$ & $\mathrm{Ku}(\eta)$ \\
		\hline
		$1$ & $-1.2065335745820$ & $1.607781034581$ & $0.29346452408$ & $0.1652429384$ \\
		\hline
		$2$ & $-1.771086807411$& $0.8131947928329$ & $0.224084203610$ & $0.0934480876$\\
		\hline
	\end{tabular}
\end{center}

\subsection{Details on the Extraction of the Full Counting Statistics} \label{SI:FCS_numerics}

In Fig.~3(a) of the main we displayed the full counting statistics (FCS) of the ASIP and ASEP for step initial conditions.
In this section we briefly sketch how these were obtained numerically. 
For this we start by considering $\mathcal{N}(t)$ for some late time $t$.
In our numerics the final time of our simulation was $t_{\mathrm{max},+} = 20 \gamma^{-1}$ for bosons ($t_{\mathrm{max},-} = 120 \gamma^{-1}$ for fermions) respectively.
The counting statistics we report are indicative of times later than $t \ge 0.2 \times t_{\mathrm{max},\sigma} $.
The $j_{0,\sigma}$ is known from analytics, but $\Gamma$ is only fixed up to a constant by dimensional analysis. 
The main numerical difficulty is therefore to find the appropriate rescaling $(\Gamma t)^{1/3}$ of the TWD. 
We solve this by considering the quantity
\begin{equation}
\hat{\eta} = \frac{\mathcal{N}(t) - j_{0,\sigma}t}{t^{\frac{1}{3}}} = \Gamma_\sigma^{\frac{1}{3}}\eta,
\end{equation}
which is just a rescaled version of $\eta$ from the TWD.
We start by considering the quantities
\begin{equation}
\Gamma_1 = \left(\frac{\langle \hat{\eta}\rangle}{\langle \eta \rangle}\right)^3
\qquad
\Gamma_2 = \left(\frac{\mathrm{Var}(\hat{\eta})}{\mathrm{Var}(\eta)}\right)^{\frac{3}{2}}
\qquad
r_3 = \frac{\mathrm{Sk}(\hat{\eta})}{\mathrm{Sk}(\eta)}
\qquad
r_4 = \frac{\mathrm{Ku}(\hat{\eta})}{\mathrm{Ku}(\eta)}.
\end{equation}
In an ideal setting we would have $\Gamma_\sigma = \Gamma_1 = \Gamma_2$ and $r_3 = r_4 = 1$.
But due to only a finite number of stochastic trajectories $n_{\mathrm{sample}} = 10^3$ and a limited time $t$, over which the FCS are taken (especially for bosons), these relations will only be fulfilled approximately.
Therefore we only obtain $\Gamma_1 \approx \Gamma_2 \approx \Gamma_\sigma$ as well as $r_1 \approx r_2 \approx 1$.
Additionally allowing for miniscule renormalization of $j_{0,\sigma}$ we obtain the following results summarized in the table below,
\begin{center}
	\begin{tabular}{|c|c|c|c|c|c|}
		\hline  
		$\sigma$ & $j_{0,\sigma} \,\mathrm{(th)}$ & $j_{0,\sigma}\,\mathrm{(num)}$ & $\bar{n}_\sigma$ & $A_\sigma^2$ & $\Gamma_\sigma/\gamma \,\mathrm{(num)}$ \\
		\hline
		$+$ & $2$ & $2.012$ & $1$ &$4 $ & $1.2 $ \\
		\hline
		$-$ & $0.25$& $0.2425$ & $\frac{1}{2}$& $\frac{1}{16} = 0.0625 $ & $0.075 $\\
		\hline
	\end{tabular}
\end{center}
where we have se $\gamma = 1$ in all simulations.
For bosons and fermions we assume different mean fillings $n_{\sigma}$ in the origin for long times.
For fermions the noiseless hydrodynamics predict a rarefaction wave with $n_0(t\rightarrow\infty) = \bar{n}_{-} = \frac{1}{2}$ in the origin, which also corresponds to the mean density.
For bosons the noiseless hydrodynamics predict long time occupations of $\bar{n}_+ = 1$.
Similar to Eq.~\eqref{eq:Asigma} we defined $A_\sigma = \bar{n}_\sigma (1+\sigma \bar{n}_\sigma)$ but with the local occupations $\bar{n}_{\sigma}$ instead of the filling $\bar{n} = \frac{1}{2}$.
This suggests that for fermions we have
\begin{equation}
    C_-  \approx 1 
\end{equation}
which indeed corresponds to the exactly know value $C_- = 1$ \cite{Johansson2000_SI} up to numerical uncertainty.
For bosons on the other side the exact value of the constant $C_+$ is not known yet, but our numerical results suggest
\begin{equation}
    C_+ \approx 0.3.
\end{equation}

\section{Supplement to Paragraph: ``Large deviation analysis for bosons"}

\subsection{Asymptotics of the TWD}
In our simplified large deviation analysis we obtained the asymptotic scaling of the FCS for the case of bosons.
In order to be self-contained we also give the known asymptotics of the TWD, which were originally obtained in \cite{Tracy93_SI} and are of the form
\begin{eqnarray}
		\mathrm{TWD}_{\beta}(\eta) \sim 
	\begin{cases}
		e^{-\frac{\beta}{24}\vert \eta\vert^3} & \eta \rightarrow -\infty \\
		e^{-\frac{2\beta}{3}\eta^{\frac{3}{2}}} & \eta \rightarrow + \infty.
	\end{cases}
\end{eqnarray}

\section{Supplement to paragraph ``Unified picture of boson and fermion transport"}

\subsection{KPZ Equation: Transformation from Transport to Interface Growth}
We start by clearly stating the KPZ equation~\eqref{eq:KPZ_transp} in the context of transport
\begin{equation} \label{eq:KPZ_SI_I}
	\partial_t h(t,x) = \underbrace{0}_{ = v_\sigma} - \underbrace{\gamma}_{ = c_\sigma} \partial_x  h(t,x)+ \gamma \sigma (\partial_x h(t,x))^2 + \nu \partial_x^2 h(t,x) + \xi(t,x), 
\end{equation}
with the initial condition
\begin{equation}
h(0,x) = \vert x \vert \theta(-x).
\end{equation}
Here, $\theta(x)$ is the Heaviside step function and the initial condition is obtained from $n(t,x) = -\partial_x h(t,x)$ and the step initial condition $n(t=0,x) = \theta(-x)$.
We will now make the transformation
\begin{equation}
h(t,x) \rightarrow h(t,x) + \frac{x}{2},
\end{equation}
which on the level of the discrete lattice simulations corresponds to shifting $h_{i+\frac{1}{2}} \rightarrow h_{i+\frac{1}{2}} + \frac{i}{2}$.
From this we obtain the KPZ equation
\begin{equation} \label{eq:SI_KPZ_corner}
\partial_t h(t,x) = \underbrace{\frac{\gamma}{2} \left( 1 + \frac{\sigma}{2}\right)}_{ = v_\sigma} - \underbrace{\gamma(1 + \sigma )}_{=c_\sigma} \partial_x h(t,x) + \sigma  \gamma (\partial_x h(t,x))^2 + \nu \partial_x^2 h(t,x) + \xi(t,x)
\end{equation}
with initial condition
\begin{equation}
h(t=0,x) = \frac{\vert x \vert }{2}.
\end{equation}

\subsection{Curvature from the Dissipationless KPZ Equation}

In the main paper the curvature of the given interface was crucial for the detailed counting statistics, which is GOE (GUE) for a flat (curved) fluctuating interface.

Here we determine the curvature of the interface from hydrodynamics in the point $x = 0$, more concretely we evolve the noiseless deterministic KPZ equation,
\begin{equation}
	\partial_t h(t,x) = v_\sigma-c_\sigma \partial_x h(t,x) +  \sigma \gamma (\partial_x h(t,x))^2 + \nu \partial_x^2 h(t,x),
\end{equation}
with $c_\sigma$ and $v_\sigma$ from Eq.~\eqref{eq:SI_KPZ_corner} with corner initial conditions $h(0,x) = \vert x \vert/2$.
Note that although we are interested in the limit $\nu \rightarrow 0$ will keep it and consider the limit later.

Defining $h(t,x) = v_\sigma t + h^\prime(t,x)$ the noiseless KPZ is mapped on a heat equation via the transformation $Z(t,x) = \exp(\sigma \frac{\gamma}{\nu}h^\prime(t,x))$ and we obtain
\begin{equation}
	\partial_t Z(t,x) = -c_\sigma \partial_x Z(t,x) + \nu \partial_x^2 Z(t,x).
\end{equation}
We solve this equation using the heat-kernel and subsequently taking the logarithm we obtain
\begin{equation}
    h(t,x) = v_\sigma t + \sigma \frac{\nu}{\gamma} 
    \log
    \left\{  \int \mathrm{d}x^\prime \frac{1}{\sqrt{4\pi \nu t}}\exp\left(-\frac{( x-c_\sigma t-x^\prime)^2}{4\nu t}+\sigma \frac{\gamma}{\nu}h(0,x^\prime)\right)\right\}.
\end{equation}
In the limit $\nu \rightarrow 0$ we obtain
\begin{equation}
    h(t,x) = v_\sigma t + 
    \underset{y}{\max} \left\{ 
    \frac{\vert y \vert}{2} - \sigma \frac{(x-c_\sigma t-y)^2}{4\gamma t}
    \right\}
\end{equation}
where we have neglected a infinitely large constant.

Solving this optimization problem we obtain for the case of surface erosion $\sigma = -1$ the height profile
\begin{equation}
	h(t,x) =
	\begin{cases}
		\frac{\gamma t}{4} + \frac{x^2}{4\gamma t} & \vert x \vert \le \gamma t \\
		\frac{\vert x \vert }{2} & \vert x \vert > \gamma t.
	\end{cases}
 \qquad
 \sigma = -1
\end{equation}
This is consistent with the hydrodynamic solutions of the noiseless Burgers equation for fermions in Sec.~\ref{SI:Nt}.

We will now consider the novel case of the surface growth process with $\sigma = 1$, which corresponds to the bosonic transport.
In this case the height profile is
\begin{equation}
	h(t,x)  = \gamma t + \frac{1}{2}\vert x - c_+ t\vert
 \qquad
 \sigma = +1
\end{equation}
In Fig.~\ref{fig:3} in the main text, these solutions of the noise-and diffusionless KPZ equation are indicated by the dashed lines.
Around the origin, we have $h(t,x)=(\gamma+\frac{c_+}{2})-\frac{x}{2}$; hence, the surface around $x=0$ is always plane (albeit tilted). 
This is in agreement with our finding that the fluctuations follow a TWD of GOE type.
Note that in the case $c_+ = 0$ the corner is growing symmetrically and despite the singularity in the origin we numerically verified that the FCS are still of the GOE type. 
Therefore we conclude that the GOE counting statistics is only attributed to the positive sign of the KPZ nonlinearity.

\end{document}